\documentclass[11pt]{article}
  
\usepackage{amsmath,epsfig}
\usepackage{array}
\begin{document}
  
\title{{
Hamiltonian solutions of the 3-body problem in (2+1)-gravity
}
}

\author{{M. Ciafaloni}\\
{\em\small Dipartimento di Fisica, Universit\`a di Firenze
and INFN, Sezione di Firenze,}\\
{\em\small 50019 Sesto Fiorentino, Italy}\\
\\
{S. Munier}\\
{\em\small Centre de Physique Th\'eorique, \'Ecole Polytechnique, CNRS,}
{\em\small 91128 Palaiseau, France}}

\maketitle

\begin{abstract}
We present a full study of the 3-body problem
in gravity in flat (2+1)-dimensional space-time,
and in the nonrelativistic limit of small
velocities.
We provide 
an explicit form of the ADM Hamiltonian in a regular coordinate
system and we set up all the ingredients for canonical quantization.
We emphasize the r\^ole of a $U(2)$ symmetry under which the Hamiltonian
is invariant
and which should generalize to a $U(N-1)$ symmetry for $N$ bodies.
This symmetry seems to stem from a braid group structure
in the operations of looping of particles around
each other, and
guarantees the single-valuedness of the Hamiltonian.
Its r\^ole for the construction 
of single-valued energy eigenfunctions is also discussed.
\end{abstract}

\section{Introduction}
The gravitational problem with or without 
matter in 2+1 dimensions 
\cite{S} has received
considerable attention in the past years 
(for reviews, see Ref.~\cite{CarlipBook,Carlip2005})
as a laboratory for a
nonperturbative treatment of gravity at the classical and quantum level.
Recently, much work has been devoted to topologically
massive 2+1-gravity in AdS spacetimes (see e.g. \cite{DC}
and references therein),
but several aspects of the more conventional problem with
matter in an open spacetime are still unsolved, and
addressed here.

In fact, despite several efforts, 
an explicit quantum mechanical treatment of $O(2,1)$
gravity with matter has been found only in the 2-body case, following
the Deser-Jackiw-'t Hooft (DJH) classical solution \cite{DJH}
and its quantization \cite{H,DJ}. 
Subsequently, progress in the treatment of the classical $N$-body
case with a regular metric has been achieved both in the first-order formalism
\cite{BCV1,BCV2} and, more implicitly, in the canonical one \cite{CMS} by using
a York-type gauge. But no conclusive work on the solutions 
of the $N$-body quantum gravity
problem is yet available for $N\geq 3$.

The purpose of the present paper is to perform further steps in the direction
of a canonical quantum treatment of the $N=3$ case. By using the simplifying
assumption of small velocities \cite{BV,BCV2}, 
we are able to provide the (single-valued)
Hamiltonian and a complete set of constants of motion in a fully explicit
form and in a regular coordinate system. Canonical quantization is then,
in principle, straightforward. However, we have not been able, so far, to
implement the monodromy condition on the energy  
eigenfunctions, and thus to construct
the canonical Hilbert space.

We use, throughout the paper, a regular coordinate system, such that
the metric be single-valued everywhere, including in the neighbourhoods
of (pointlike) external matter. On the other hand, it is known \cite{S}
that space-time is flat outside matter sources, so that
one can use instead Cartesian coordinates -- characterized by various
deficit angles -- in which the
conjugate momenta to the particles'
positions are constants of motion.
The mapping from regular coordinates in a York-type gauge to
Cartesian coordinates was constructed in Ref.~\cite{BCV2}
thus providing an explicit expression of the constants of motion
both for $N=2$ and in the nonrelativistic $N=3$ case.

The canonical formalism was then set up in Ref.~\cite{CMS}
by providing the form of the Hamiltonian for $N=2$
and its implicit definition for $N\geq 3$. Its general interpretation
is that of the scale factor in the asymptotic Liouville field
occuring in the ADM parametrization
of the metric. The external masses act as sources of the Liouville
field, together with the so-called apparent singularities \cite{Y}
of the problem. Therefore, the asymptotic scale factor is a function
of the particle masses and coordinates which can in principle be computed.

Our first task here is to provide the explicit form of the Hamiltonian
for the 3-body case, on the basis of its definition in Ref.~\cite{CMS}.
We are able to do that in the small-velocity limit, where we show that
the Hamiltonian is simply related to a sum
of squared moduli of two properly-defined relative momenta
(called $P_3$ and $P_2$ in the paper), whose expressions in terms of
regular particle coordinates and momenta are explicitly given
in Sec.~\ref{sec4}.
Since such expressions carry branch cuts, there are nontrivial
monodromy transformations for $P_3$ and $P_2$ when the particle positions
(of particles $\#2$ and $\#3$, say) turn around each other. Nevertheless,
the Hamiltonian is left invariant (and is thus single-valued)
because it possesses a $U(2)$ symmetry and, moreover, 
the pair $(P_3,P_2)$
turns out to transform as a $U(2)$ spinor.

The $U(2)$ invariance of our problem has thus an important r\^ole
in assessing the monodromy of the Hamiltonian, and the exchange
symmetry of its equation of motion. We feel it should have a r\^ole
in quantization as well, because it regulates the degeneracy of the
wave functions and their monodromy properties.

We are thus able to provide explicit solutions of the classical
Hamilton equations and to express them in terms of Cartesian
coordinates and momenta. Such solutions agree with those found
in Ref.~\cite{BCV2} in the first-order formalism.
Given such explicit understanding of the Hamiltonian structure, canonical
quantization is in principle straightforward, 
and is here formulated by using a 
proper ordering prescription for the expression of the Hamiltonian 
in terms of regular coordinates and momenta.
However, while the Hamiltonian is single valued, its eigenfunctions
-- characterized by a large degeneracy -- are generically multiple-valued
under the braiding of the 2 and 3 labels, for instance.
We have some ideas on how to possibly construct monodromic
eigenstates, but we have not found a successful procedure yet.

After summarizing previous work on the canonical formalism
and describing how to calculate the Hamiltonian in the nonrelativistic limit
in Sec.~\ref{sec2}, we provide its explicit form for $N=3$ 
in Sec.~\ref{sec4}, where we discuss its $U(2)$ symmetry also.
The equations of motion and their classical solutions are given
in Sec.~\ref{sec5}, 
the comparison with previous calculations
in a different formalism is done in Sec.~\ref{sec6}
while in Sec.~\ref{sec7}, we sketch
the quantization of the problem. We summarize
our results and outline a few ideas which may prove useful
in order to construct the quantum Hilbert space in the final section.
Some technical details are discussed in the appendix.

\section{\label{sec2}Hamiltonian formulation and nonrelativistic limit}

We want to describe the motion of $N$ pointlike massive 
particles in a 2+1 dimensional open universe in a Hamiltonian
formalism, in order to be able to perform canonical quantization.
The complete formalism being available in several places
in the literature \cite{MS,CMS} (see also \cite{BCV1,W}), 
we shall not enter the details of the
derivation.
We provide the essential formulae in the general
case in the first subsection,
and specialize to the nonrelativistic limit in the next one.

\subsection{General Hamiltonian for the motion of $N$ pointlike massive
particles}

Let us start with the action in the
Arnowitt-Deser-Misner (ADM) formulation~\cite{ADM1959,ADM}.
Parametrizing the generic line element $ds$ in the standard form
\begin{equation}
ds^2=-N^2 dt^2+h_{ij}(dx^i+N^i dt)(dx^j+N^j dt),
\end{equation}
the action reads
\begin{equation}
S=\int dt\left\{
\sum_n p_{ni}\dot q_n^i+\int
d^2x(\Pi^{ij}\dot h_{ij}-N^i{\cal H}_i-N{\cal H})
\right\}
+S_B,
\label{action}
\end{equation}
where $q_n$ is the position of particle number $n$ ($p_n$ is
the conjugate momentum), $\Pi$ is the canonical
momentum conjugate to the spatial metric $h$, and
\begin{equation}
\begin{split}
{\cal H}_i&=-2\sqrt{h}\nabla_j\frac{\Pi^j_{\ i}}{\sqrt{h}}
-\sum_n \delta^2(x-q_n)p_{ni}\\
{\cal H}  &=\frac{2\kappa^2}{\sqrt{h}}
\left[\Pi^i_{\ j}\Pi^j_{\ i}-(\Pi^i_{\ i})^2\right]
-\frac{\sqrt{h}{\cal R}}{2\kappa^2}
+\sum_n\delta^2(x-q_n)\sqrt{m_n^2+h^{ij}p_{ni}p_{nj}}.
\end{split}
\label{eq:constraints}
\end{equation}
${\cal R}$ is the intrinsic curvature of the 2-spacelike slice and
$\nabla$
is the covariant derivative compatible with the 2-dimensional metric
$h_{ij}$.
$\kappa^2$ is related to Newton's constant
through $\kappa^2=8\pi G$, and
has dimensions of an inverse mass in 2-dimensional space. 
It was set to $\frac12$ in
Ref.~\cite{MS,CMS}, and to 1 in Ref.~\cite{DJH,BCV1,BCV2}.
We refer the reader to~\cite{HH,MS} for a complete expression
including the boundary terms $S_B$.

We can check that the variation of 
the action in Eq.~(\ref{action})
with respect to the metric components $N$, $N^i$, $h_{ij}$ and 
to the
momentum $\Pi_{ij}$ yields the Einstein equations in a first-order
form. 
In particular, one gets the equations ${\cal H}=0$ and ${\cal H}_i=0$
which are the so-called Hamiltonian and
momentum constraints respectively, and which
we shall analyze below.  
The Hamiltonian for the particles' motion is obtained
after all variables have been expressed as a function of $p_n$ and
$q_n$.

We choose the
York instantaneous gauge in which the
intrinsic curvature $K$ is zero everywhere, 
which has proved useful in this context.
Introducing complex notation
$z\equiv x^1+ix^2$, $z_n\equiv q_n^1+iq_n^2$,
and $p_n\equiv({p_n^1-ip_n^2})/{2}$,
we impose the further gauge-fixing conditions: $h_{zz}=0$ and 
$h_{\bar z\bar z}=0$. The spatial line element then
takes the conformally flat form
\begin{equation}
dl^2=e^{2\sigma}|dz|^2\ \ (\mbox{or, equivalently,}\ \
h_{ij}=e^{2\sigma}\delta_{ij}). 
\label{lineelement}
\end{equation}
This equation defines the
field $\sigma$.

The momentum and Hamiltonian constraints can now be solved
for
$\Pi$ and $\sigma$. 
The momentum constraint
${\cal H}_i=0$ determines the components
of the energy-momentum tensor:
\begin{equation}
\Pi\equiv\Pi^{\bar z}_{\ z}=-\frac{1}{2\pi}\sum_n \frac{p_n}{z-z_n}
\end{equation}
(up to an entire function of the $z$ variable which has to be
set to zero to ensure good asymptotic properties of the metric, see
Ref.~\cite{MS}).
We will work in a center-of-mass frame in which $\sum_n p_n=0$. In
this case, $\Pi$ is the ratio of two polynomials of respective degree
$N-2$ and $N$. Let us denote by $z_A$ the zeros of the polynomial in
the numerator. They are functions of the canonical variables $z_n$ and
$p_n$. We write
\begin{equation}
\Pi=-\frac{D}{2\pi}\frac{\prod_A(z-z_A)}{\prod_n(z-z_n)},
\label{eq:Pi}
\end{equation}
where the normalization $D$ is the so-called dilation factor 
\begin{equation}
D\equiv\sum_n z_n p_n,
\label{dilation0}
\end{equation}
whose imaginary part is half of the total angular momentum
of the system of particles.

The Hamiltonian constraint ${\cal H}=0$
reduces to the nonlinear equation
\begin{equation}
\Delta(2\sigma)=-2|2\kappa^2\Pi|^2 
e^{-2\sigma}-4\pi\sum_n\mu_n\delta^2(z-z_n),
\label{liouville}
\end{equation}
where 
$\Delta\equiv\partial_x^2+\partial_y^2=4\partial_z\partial_{\bar z}$ 
and where
we have introduced
the dimensionless masses
$\mu_n\equiv\kappa^2 m_n/{2\pi}$.
Defining
\begin{equation}
e^{-2\tilde\sigma}\equiv 2|2\kappa^2\Pi|^2 e^{-2\sigma},
\label{eq:deftildesigma}
\end{equation}
the new field $\tilde\sigma$ obeys a Liouville equation:
\begin{equation}
\Delta(2\tilde\sigma)=-e^{-2\tilde\sigma}
-4\pi\sum_n(\mu_n-1)\delta^2(z-z_n)
-4\pi\sum_A\delta^2(z-z_A).
\label{liouvilletilde}
\end{equation}
There are $N$ sources located at the particles' positions $z_n$.
The remaining sources $z_A$ are so-called
apparent singularities:
They stem from the zeros of the momentum tensor $\Pi$.
Their number is $N-2$ in the center-of-mass frame.

The Euler characteristic of the
2-surface is an important parameter.
It reads
\begin{equation}
2\pi\mu=\frac12\int d^2z\sqrt{h}{\cal R}=-\frac12\int d^2z\,\Delta(2\sigma)=
\lim_{r\rightarrow\infty}-\frac12\oint_r \vec\nabla(2\sigma)\cdot d\vec n.
\label{euler}
\end{equation}
The notation for the integration element is 
$d^2z\equiv dx dy=dz d\bar z/(2i)$.
The last equality, obtained from the Gauss law on a circle of
radius $r$ with exterior normal $\vec n$, provides the asymptotic
behavior of $\sigma$ in the form
\begin{equation}
e^{2\sigma}\underset{|z|\gg|\lambda|}{\sim}
\left|\frac{z}{\lambda}\right|^{-2\mu},
\label{asymptoticsigma}
\end{equation}
where $\lambda$ is a distance scale. 

The parameter $\mu$ is 
a constant of motion
and is interpreted as the (rescaled) total mass of the universe.
If all particles are static, there is no interaction energy,
since in 2+1 dimensional gravity, spacetime is
flat outside the sources, and thus there are no local interactions.
Hence in the static 
case, the total mass of the universe is the
sum of the masses of all particles: $\mu=\sum_n\mu_n$.

It was found in Ref.~\cite{CMS} that 
the physical Hamiltonian is related to
the logarithm of the scale $\lambda$
of the spatial distances, namely
\begin{equation}
{H}=\frac{\mu}{2\kappa^2}\ln |\lambda|^{2}
\label{eq:Hscale}
\end{equation}
up to a constant term,
where $\lambda$
is a function of the
canonical variables which are the particle
momenta $p_n$ and positions $z_n$. 
To give sense to 
Eq.~(\ref{eq:Hscale}), one has
to make $\lambda$ dimensionless by dividing out a basic length
scale (which does not appear to be fundamental, but
rather related to the initial conditions.)
In the following, all length variables
($\lambda$, $z$, $z_n$\dots) 
will be considered dimensionless.

The Hamiltonian for the 2-body problem 
can be computed exactly from Eq.~(\ref{eq:Hscale}).
However, for larger values of $N$, 
the exact resolution of the Hamiltonian
constraint~(\ref{liouville}) is already 
a formidable task~\cite{YMO}.
Therefore, in this paper,
we shall stick to the nonrelativistic approximation which was proposed
in Ref.~\cite{BCV2}, and which we shall now derive in the
Hamiltonian formulation.

\subsection{\label{sec3}Nonrelativistic limit}

To characterize the quasi-static approximation, 
we have at our disposal
the $N+1$ parameters $\mu$ and $\mu_n$. 
When the particles are at
rest, the difference between the total mass of the universe
and the sum of the masses of the particles
\begin{equation}
\varepsilon\equiv\mu-\sum_n\mu_n
\end{equation} 
is zero. So in the following, we shall always restrict
ourselves to the lowest order in $\varepsilon$.

Let us start from the definition of the Euler
characteristic~(\ref{euler}).
Inserting Eq.~(\ref{liouville}) in Eq.~(\ref{euler}),
we obtain the following relation for $\tilde\sigma$  
defined in Eq.~(\ref{eq:deftildesigma}):
\begin{equation}
\frac{1}{2\pi}\int d^2z e^{-2\tilde\sigma}=\varepsilon.
\label{eulerbis}
\end{equation}
An analysis of Eq.~(\ref{liouvilletilde}) shows that on the particle
singularities, the term
$e^{-2\tilde\sigma}$ has to behave like
$|z-z_n|^{2(\mu_n-1)}$. (This is consistent with the fact
that $e^{-2\sigma}$ vanish for $z=z_n$, as long as the masses $\mu_n$
are positive).
In the right-hand side of Eq.~(\ref{liouvilletilde}), 
the $\delta$-functions then
dominate at the particle singularities. Choosing a
regularization,
we see that from Eq.~(\ref{eulerbis}), the term $e^{-2\tilde\sigma}$
is of order $\varepsilon$. The scale $\varepsilon$ disappears
when one takes the Laplacian of its logarithm as in the left-hand side.
So setting $\varepsilon$ to zero, we see that the nonlinear term
drops out and the Liouville equation for $\tilde\sigma$ boils
down to a Poisson equation. 
The solution of the latter reads
\begin{equation}
e^{-2\tilde\sigma(z)}=|K|^2\prod_{n=1}^N|z-z_n|^{2\mu_n-2}
\prod_{A=1}^{N-2}|z-z_A|^{2},
\label{sigmatildeprod}
\end{equation}
where $K$ is a complex number
independent of $z$. We note that this solution
is consistent with the one found in Ref.~\cite{BCV2}.
There, the analysis was performed in a first-order formalism,
using the scale of the
Cartesian particle velocities as a small parameter.
We shall make the comparison sharper later on.

To obtain
the Hamiltonian, we only
need the large-distance behavior of $e^{-2\sigma}$
which can be deduced from the behavior of $e^{-2\tilde\sigma}$
and of $\Pi$ using Eq.~(\ref{eq:deftildesigma}).
First, from Eq.~(\ref{sigmatildeprod}),
\begin{equation}
e^{-2\tilde\sigma}\underset{|z|\rightarrow\infty}{\sim}
 |K|^2|z^2|^{\sum\mu_n-2}.
\end{equation}
Second, from Eq.~(\ref{eq:Pi}),
the momentum $\Pi$ behaves like $D/z^2$,
where $D$ is defined in Eq.~(\ref{dilation0}). 
Identifying the asymptotic behavior of $e^{-2\sigma}$
just found
with Eq.~(\ref{asymptoticsigma})
at the
lowest order in $\varepsilon$, we 
get an expression for $|\lambda|^2$ from which,
with the help of Eq.~(\ref{eq:Hscale}), we
deduce
the following formula
for the Hamiltonian:
\begin{equation}
{H}=\frac{1}{2\kappa^2}\ln\frac{|2 \kappa^2 D|^2}{|K|^2},
\label{hamiltonian0}
\end{equation}
up to an irrelevant constant. The parameter $|K|^2$ can be determined
as a function of the canonical variables and of the constant
$\varepsilon$ using
Eq.~(\ref{eulerbis}).
So our task is now 
to integrate Eq.~(\ref{sigmatildeprod})
over the whole complex plane. An appropriate change of variable
enables us to cast the integral in the form
\begin{equation}
\varepsilon=\frac{1}{4\pi}|K|^2|z_{21}|^{2\mu-2}
\int d^2\xi |\xi|^{2\mu_1-2}|1-\xi|^{2\mu_2-2}|\zeta_3-\xi|^{2\mu_3-2}
\cdots|\zeta_N-\xi|^{2\mu_N-2}
\prod_A|\zeta_A-\xi|^2
\label{epsilongeneral}
\end{equation}
where we have defined the new variable 
$\zeta_{n/A}\equiv(z_{n/A}-z_1)/(z_2-z_1)$, $z_{ij}=z_i-z_j$
and $\mu=\sum_{n=1}^N\mu_n$
consistently with the quasi-static approximation.

It is easy to check that these formulae allow to recover,
up to an additive constant, the Hamiltonian of the
2-body problem, which was written down
by several groups \cite{DJH,H,DJ,BCV1,BCV2,CMS}:
\begin{equation}
{H}=\frac{1}{2\kappa^2}
\left(
\ln |2\kappa^2 p|^2+\mu\ln|z|^2
\right),
\label{hamiltonian2bodies}
\end{equation}
where $p$ is the momentum $p_2$ of particle \#2 (keeping
in mind that $p_1+p_2=0$), 
$z\equiv z_{21}=z_2-z_1$.

\section{\label{sec4}Explicit Hamiltonian for the three-body problem}

So far, we have obtained a general formula for the Hamiltonian
in the nonrelativistic limit as a function of the canonical
positions and momenta of the particles,
described in a regular coordinate
system by the variables $z_n$ and $p_n$ respectively.
It is given in Eq.~(\ref{hamiltonian0}), with $D$ 
defined in Eq.~(\ref{dilation0})
and $|K|^2$ obtained from the evaluation of the
integral in~(\ref{epsilongeneral}).
In the present and the following sections, we shall 
specialize to
$N=3$ bodies. In this case, the nonrelativistic Hamiltonian
may still be expressed in terms of known functions and, therefore,
may be studied completely.

Interesting new features appear with respect
to the two-body case.
In particular, the Hamiltonian possesses a $U(2)$ invariance,
related to a braid group structure of
the particle exchanges and loopings.

\subsection{Explicit calculation}

Let us first gather the ingredients for the computation
of the Hamiltonian~(\ref{hamiltonian0}), namely
the expressions of $|K|^2$ and of $D$.

We specialize Eq.~(\ref{epsilongeneral}) to the case of 3 bodies.
The rescaled position variables of particle \#3 and of the apparent
singularity respectively read
\begin{equation}
\zeta\equiv\zeta_3=\frac{z_{31}}{z_{21}},\ 
\zeta_A=\frac{(p_2+p_3)\zeta}{p_2+{p_3}\zeta}.
\label{eq:zetaA}
\end{equation}
We choose
the origin of the frame at the position of
particle~\#1
in such a way that $z_1=0$.
Then Eq.~(\ref{epsilongeneral}) reads
\begin{equation}
{\varepsilon}=\frac{1}{4\pi}{|K|^2}|z_2|^{2\mu-2}\int d^2 \xi
|\xi|^{2\mu_1-2}|1-\xi|^{2\mu_2-2}
|\zeta-\xi|^{2\mu_3-2}
|\zeta_A-\xi|^2,
\label{secondbis}
\end{equation}
where $\mu=\mu_1+\mu_2+\mu_3$ in the right-hand side
of this equation.
The dilation operator defined in Eq.~(\ref{dilation0}) is
\begin{equation}
D=z_2(p_2+p_3\zeta).
\label{D}
\end{equation}

We now compute the integral that defines $\varepsilon$ 
(Eq.~(\ref{secondbis})).
To this aim, we first expand
\begin{equation}
|\xi-\zeta_A|^2=|\xi-\zeta|^2
+(\xi-\zeta)(\bar\zeta-\bar\zeta_A)
+(\bar \xi-\bar\zeta)(\zeta-\zeta_A)
+|\zeta-\zeta_A|^2
\end{equation}
in order to be able to cast $\varepsilon$ 
in the form of a sum
of standard integrals:
\begin{equation}
\varepsilon=|K|^2|z_2|^{2\mu-2}\left[J_{11}
+(\bar\zeta_A-\bar\zeta)J_{10}
+(\zeta_A-\zeta)J_{01}
+|\zeta_A-\zeta|^2 J_{00}
\right],
\label{epsilonsum}
\end{equation}
where
\begin{equation}
J_{\delta\tilde\delta}=\int d^2 \xi |\xi|^{2(\mu_1-1)}|1-\xi|^{2(\mu_2-1)}
(\zeta-\xi)^{\mu_3-1+\delta}
(\bar\zeta-\bar \xi)^{\mu_3-1+\tilde\delta}.
\end{equation}
This kind of integrals appears in conformal field theory 
and were computed e.g. in Refs.~\cite{DF,GN}.
The $J$-functions 
can be expressed
with the help of hypergeometric functions:
\begin{multline}
{J_{\delta\tilde\delta}}=
\zeta^{\mu_1+\mu_3+\delta-1}\bar\zeta^{\mu_1+\mu_3+\tilde\delta-1}
B_{\mu_1,\mu_3+\delta}
F(\mu_1,1-\mu_2,\mu_1+\mu_3+\delta,\zeta)\\
\times B_{\mu_1,\mu_3+\tilde\delta}
F(\mu_1,1-\mu_2,\mu_1+\mu_3+\tilde\delta,\bar\zeta)
\frac{s_1 s_3}
{s_{13}}\\
+(-1)^{\delta-\tilde\delta}
B_{\mu_2,\mu_1+\mu_3+\delta-1}F(1-\mu_3-\delta,2-\mu_1-\mu_2-\mu_3-\delta,
2-\mu_1-\mu_3-\delta,\zeta)\\
\times B_{\mu_2,\mu_1+\mu_3+\tilde\delta-1}
F(1-\mu_3-\tilde\delta,2-\mu_1-\mu_2-\mu_3-\tilde\delta,
2-\mu_1-\mu_3-\tilde\delta,\bar\zeta)
\frac{s_2 s_{13}}{s_{123}},
\label{J0}
\end{multline}
where 
$s_i=\sin\pi\mu_i$,
$s_{ij\cdots}=\sin\pi(\mu_i+\mu_j+\cdots)$
and $B_{\alpha,\beta}$ is the standard Euler beta function.
Note that since we are working at 
the lowest order in $\varepsilon$, 
we may always replace
$\sum \mu_i$ by the total mass $\mu$ 
in each term that appears in the r.h.s. of 
Eq.~(\ref{epsilonsum}).
We define the four following functions:
\begin{equation}
\begin{array}{ll}
f_{3a}=B_{\mu_1,\mu_3}\zeta^{\mu_1+\mu_3-1}F(\mu_1,1-\mu_2,\mu_1+\mu_3,\zeta),
& f_{3b}=f_{3a}|_{\mu_3\rightarrow 1+\mu_3},\\
f_{2a}=B_{\mu_1+\mu_3-1,\mu_2}F(1-\mu_3,2-\mu,2-\mu_1-\mu_3,\zeta),
&f_{2b}=f_{2a}|_{\mu_3\rightarrow 1+\mu_3}.
\end{array}
\label{f}
\end{equation}
The functions
$f_{3a}$ and $f_{2a}$ of the variable $\zeta$
are two independent solutions of the hypergeometric 
equation
\begin{equation}
\zeta(1-\zeta)f_{ia}^{\prime\prime}+[2-\mu_1-\mu_3-(3-\mu_3-\mu)\zeta] 
f_{ia}^\prime
-(2-\mu)(1-\mu_3)f_{ia}=0
\label{hyper}
\end{equation}
and, similarly, $f_{3b}$ and $f_{2b}$ solve the hypergeometric equation
obtained from the previous one after having performed 
the shift $\mu_3\rightarrow \mu_3+1$.

After some algebra, 
involving in particular the trigonometric identity
\begin{equation}
\sin\pi\alpha\,
\sin\pi(\alpha+\beta+\gamma)
=\sin\pi(\alpha+\beta)
\sin\pi(\alpha+\gamma)
-\sin\pi\beta\,
\sin\pi\gamma
\label{idtrig}
\end{equation}
with $\alpha=\mu_1$, $\beta=\mu_2$ and $\gamma=\mu_3$,
we arrive at an expression of ${\varepsilon}/{|K|^2}$ in terms
of the $f$'s:
\begin{equation}
\frac{4\varepsilon}{|K|^2}=|z_2|^{2\mu-2}\left[
 N_3|(\zeta_A-\zeta)f_{3a}+f_{3b}|^2
+N_2|(\zeta_A-\zeta)f_{2a}-f_{2b}|^2
\right],
\label{finalepsilon}
\end{equation}
where
\begin{equation}
N_2\equiv\frac{s_2 s_{13}}{\pi s_{123}},\ 
N_3\equiv\frac{s_1 s_3}{\pi s_{13}}.
\label{N}
\end{equation}
The Hamiltonian~(\ref{hamiltonian0}) then reads
\begin{equation}
H=\frac{1}{2\kappa^2}
\bigg\{
\ln\frac{|2\kappa^2 D|^2}{4\varepsilon}+(\mu-1)\ln |z_2|^2+
\ln\left[
N_3|(\zeta_A-\zeta)f_{3a}+f_{3b}|^2
+N_2|(\zeta_A-\zeta)f_{2a}-f_{2b}|^2
\right]
\bigg\}.
\label{H0}
\end{equation}
Introducing some more notations, 
whose physical interpretation will
be given below, the Hamiltonian~(\ref{H0}) may be rewritten in a
compact form
\begin{equation}
H=\frac{1}{2\kappa^2}
\ln \frac{(2\kappa^2)^2(|P_2|^2+|P_3|^2)}{4\varepsilon},
\label{H}
\end{equation}
where we have defined
\begin{equation}
P_2\equiv Dz_2^{\mu-1}\sqrt{N_2}\left[f_{2b}-(\zeta_A-\zeta)f_{2a}\right]\ \
\mbox{and}\ \
P_3\equiv Dz_2^{\mu-1}\sqrt{N_3}\left[f_{3b}+(\zeta_A-\zeta)f_{3a}\right]
\label{P0}
\end{equation}
or, written in terms of the canonical variables $z_2,\zeta$ 
and $p_2,p_3$ with the help of Eqs.~(\ref{eq:zetaA}) and~(\ref{D}),
\begin{equation}
\begin{split}
P_2(z_2,\zeta,p_2,p_3)=z_2^\mu F_2(\zeta,p_2,p_3) ,\ \
F_2&=\alpha(\zeta)p_2+\beta(\zeta)p_3\\
P_3(z_2,\zeta,p_2,p_3)=z_2^\mu F_3(\zeta,p_2,p_3) ,\ \
F_3&=\gamma(\zeta)p_2+\delta(\zeta)p_3,
\end{split}
\label{P}
\end{equation}
the coefficients $\alpha$,$\beta$,$\gamma$,$\delta$
in front of the momenta being defined as
\begin{equation}
\begin{split}
\alpha=\sqrt{N_2}f_{2b},\
\beta&=\sqrt{N_2}\zeta(f_{2b}-(1-\zeta)f_{2a}),\\
\gamma=\sqrt{N_3}f_{3b},\
\delta&=\sqrt{N_3}\zeta(f_{3b}+(1-\zeta)f_{3a}).
\end{split}
\label{greeks}
\end{equation}
This set of notations will help to write the calculation
of Sec.~\ref{sec5} in a simpler way.

\subsection{\label{sec:interpretation}Interpretation of 
$P_3$ and $P_2$ in the small mass limit}

Before embarking with the study of the Hamiltonian~(\ref{H}), 
we wish to try and give an interpretation of
its expression.
It is quite straightforward in the small mass limit
$\mu_n\ll 1$ 
for $n=1,2,3$
in which gravity effects vanish.
Let us write the expression of $P_2$ and $P_3$ in
Eq.~(\ref{P}) in this limit.
We start with the $f$'s defined in Eq.~(\ref{f}): At the 
lowest order in all the $\mu_n$, they boil down to
\begin{equation}
f_{2a}\simeq \frac{\mu}{\mu_{13}\mu_2}\frac{1}{1-\zeta},\
f_{2b}\simeq \frac{\mu}{\mu_{13}\mu_2},\
f_{3a}\simeq \frac{\mu_{13}}{\mu_1\mu_3}\frac{1}{\zeta}
+\frac{1}{\mu_3}\frac{1}{1-\zeta},\
f_{3b}\simeq \frac{1}{\mu_1},
\end{equation}
and it follows that
\begin{equation}
P_2\simeq \frac{1}{\sqrt{N_2}}
\frac{\mu_2\mu_{13}}{\mu}
\left(\frac{p_2}{\mu_2}
-\frac{p_1+p_3}{\mu_{13}}\right),\
P_3\simeq \frac{1}{\sqrt{N_3}}
\frac{\mu_1\mu_3}{\mu_{13}}
\left(\frac{p_3}{\mu_3}
-\frac{p_1}{\mu_{1}}\right).
\label{Plimit}
\end{equation}
Thus we see that
$P_3$ is the relative momentum of particles 
\#1 and \#3 with the normalization
$\sqrt{N_3}\simeq \sqrt{\mu_1\mu_3/\mu_{13}}$, and $P_2$
is the relative momentum of particle \#2 
with respect to the (13) subsystem, with
the normalization $\sqrt{N_2}\simeq \sqrt{\mu_2\mu_{13}/\mu}$.
In the absence of interaction, 
the nonrelativistic kinetic energy of
3 pointlike particles
reads
\begin{equation}
E=2\kappa^2
\left(
\frac{\vec p_1^2}{2\mu_1}
+\frac{\vec p_2^2}{2\mu_2}+\frac{\vec p_3^2}{2\mu_3}
\right)
=
4\kappa^2\left(\frac{|p_1|^2}{\mu_1}
+\frac{|p_2|^2}{\mu_2}+\frac{|p_3|^2}{\mu_3}\right).
\end{equation}
With the help of the expressions for $P_2$ and $P_3$
in the small mass limit
given by Eq.~(\ref{Plimit}),
it can be rewritten as
\begin{equation}
E=4\kappa^2\left(|P_2|^2+|P_3|^2\right).
\label{eq:E}
\end{equation}
Note that this kinetic energy does not coincide with the
Hamiltonian in Eq.~(\ref{H}), which reads
\begin{equation}
H=\frac{1}{2\kappa^2}\ln \frac{\kappa^2 E}{4\varepsilon}.
\label{eq:Hdiffkinetic}
\end{equation}
This is actually
related to the time gauge which was chosen, and is better 
discussed in Sec.~\ref{sec:Z}.

\subsection{\label{sec:monobraids}
Monodromy properties, $U(2)$ symmetry and braids}

We want to check that the Hamiltonian~(\ref{H0}) is well-defined.
It is necessary that it be single-valued, i.e. invariant
under the loopings 
of $\zeta$ around the branch points at $0$, $1$ and $\infty$,
and invariant under the exchange of the labels of the
particles.

\subsubsection{\label{sec:mono}Monodromies and $U(2)$ symmetry}

Let us introduce the two objects
\begin{equation}
\sigma_b=\left(
\begin{matrix}
\sqrt{N_3}f_{3b}\\
\sqrt{N_2} f_{2b}
\end{matrix}
\right)\ \text{and}\
\sigma_a=\left(
\begin{matrix}
\phantom{-}\sqrt{N_3}f_{3a}\\
-\sqrt{N_2} f_{2a}
\end{matrix}
\right)
\label{defspinors}
\end{equation}
in such a way that the momenta $(P_3,P_2)$ 
defined in Eq.~(\ref{P0}) can
be conveniently rewritten as
\begin{equation}
P\equiv\left(\begin{matrix}P_3\\P_2\end{matrix}\right)
=D z_2^{\mu-1}\left[\sigma_b+(\zeta_A-\zeta)\sigma_a\right].
\label{eq:Pwithsigma}
\end{equation}
We compute the monodromy matrices $M_{32}$ and $M_{31}$
which correspond to
the loopings of particle~\#3 around particle~\#2,
and \#3 around~\#1 respectively.
These
transformations amount to
substituting
$\zeta-1\rightarrow e^{2i\pi}(\zeta-1)$
and
$\zeta\rightarrow e^{2i\pi}\zeta$.
We find that $\sigma_a$ and $\sigma_b$ transform
according to the matrix 
(see Appendix~\ref{appB} for the details)
\begin{equation}
M_{31}=e^{i\pi(\mu_1+\mu_3)}\left(
\begin{matrix}
e^{i\pi(\mu_1+\mu_3)} & 0\\
0 & e^{-i\pi(\mu_1+\mu_3)}
\end{matrix}\right)
\label{M0}
\end{equation}
for the looping of particle \#3 around particle \#1.
As for the looping of particle \#3 around particle \#2,
the corresponding transformation reads
\begin{equation}
M_{32}=e^{i\pi(\mu_2+\mu_3)}
\left(
\begin{matrix}
\phantom{-}a_{32} & b_{32}\\
-\bar b_{32} & \bar a_{32}
\end{matrix}
\right)
\ \ \mbox{with}\ \
\begin{cases}
a_{32}=\cos \pi(\mu_2+\mu_3)+i\sin\pi(\mu_2+\mu_3)\,\cos\alpha_{32},\\
b_{32}=i\sin\pi(\mu_2+\mu_3)\,\sin\alpha_{32},
\end{cases}
\label{M1}
\end{equation}
and where the angle $\alpha_{32}$ has been defined as
\begin{equation}
\cos\alpha_{32}=\frac{s_1s_2-s_3s_{123}}{s_{13}s_{23}},\
\sin\alpha_{32}=-\frac{2\sqrt{s_1s_2s_3s_{123}}}{s_{13}s_{23}}.
\end{equation}
The matrices $\tilde M_{31}=M_{31} e^{-i\pi(\mu_1+\mu_3)}$ and 
$\tilde M_{32}=M_{32} e^{-i\pi(\mu_2+\mu_3)} $ are $SU(2)$ transformations, 
so that the monodromy group generated by $M_{31}$ 
and $M_{32}$ is actually a subgroup of $U(2)$.

The object
$(\zeta_A-\zeta)\sigma_a$ also transforms according to 
$M_{31}$ and $M_{32}$, and
the same holds for the sum $(\zeta_A-\zeta)\sigma_a+\sigma_b$.
This implies that the norm of this $SU(2)$ spinor, which reads
\begin{equation}
N_3|(\zeta_A-\zeta)f_{3a}+f_{3b}|^2
+{N_2}|(\zeta_A-\zeta)f_{2a}-f_{2b}|^2
\end{equation}
is stable under the monodromy transformations.
Consequently, the Hamiltonian~(\ref{H0}) is single valued
under all possible 
loopings of particle \#3 around the two other particles.

We have not considered the looping of particle \#2 around
particle \#1.
The corresponding transformation of the spinors
can be computed:
The best is to
exchange \#3 and \#2 in such a way that
the looping of particle \#2 around \#1
be the substitution $\zeta\rightarrow e^{2i\pi}\zeta$,
and eventually, to go back to the initial coordinates
by applying the inverse exchange transformation.
Therefore, we postpone this discussion
after we have studied how the momenta change when
the particles are relabeled.

\subsubsection{Relabeling symmetry}

Exchanging the labels of
the particles yields nontrivial transformations
of the spinor $(P_3,P_2)$.
We now check that the Hamiltonian is invariant under
these transformations.

We first perform the exchange
$\mu_1\leftrightarrow\mu_3$ and
$(z_1,p_1)\leftrightarrow (z_3,p_3)$
in the expression of the $P$'s in Eq.~(\ref{P}).
Since the details are lengthy, we defer the full calculation to
Appendix~\ref{appB}. We find that the 
transformation of the spinor $(P_3,P_2)$
can be written as the multiplication by the diagonal matrix
\begin{equation}
\tau_{31}^{\mu_3\mu_1}
=
e^{i\pi(\mu_{1}+\mu_3)/2}
\left(\begin{matrix}
-e^{i\pi(\mu_{1}+\mu_3)/2} &0\\
0&e^{-i\pi(\mu_{1}+\mu_3)/2}
\end{matrix}\right).
\label{eq:P31tilde}
\end{equation}
In the same way, the exchange of the labels of
particles \#3 and \#2 
is represented by the multiplication by the matrix
\begin{equation}
\tau_{32}^{\mu_3\mu_2(\mu_1)}
\equiv
\frac{e^{i\pi(\mu_2+\mu_3)/2}}{\sqrt{s_{12}s_{13}}}
\left(\begin{matrix}
e^{-i\pi(\mu_1+\mu_2/2+\mu_3/2)}\sqrt{{s_3s_2}}&
e^{i\pi(\mu_3-\mu_2)/2}\sqrt{{s_1s_{123}}}\\
e^{i\pi(\mu_2-\mu_3)/2}\sqrt{{s_{1}s_{123}}}&
-e^{i\pi(\mu_1+\mu_2/2+\mu_3/2)}\sqrt{{s_2s_{3}}}
\end{matrix}\right).
\label{eq:P32tilde}
\end{equation}
We have written 
the two matrices
as the product of a $U(1)$ phase
and a $U(2)$ matrix of determinant $-1$.
(This is clear in the former case, and can be easily checked
using the trigonometric identity~(\ref{idtrig}) in the latter case.)
It is then obvious that
the spinorial norm
$|P_2|^2+|P_3|^2$ is left invariant
by these transformations, and since $\varepsilon$
is trivially invariant,
the same is true for the full Hamiltonian~(\ref{H}).

We check that
the monodromy transformations
are the ``squared'' of the relabeling transformations. 
More precisely,
\begin{equation}
\tau_{31}^{\mu_1\mu_3}\tau_{31}^{\mu_3\mu_1}=M_{31}\ \
\mbox{and}\ \
\tau_{32}^{\mu_2\mu_3(\mu_1)}\tau_{32}^{\mu_3\mu_2(\mu_1)}=M_{32}.
\end{equation}

We are now in a position to easily get the last monodromy
transformation which we left uncomputed
in the last subsection, namely the action of
the looping of particle \#2 around particle \#1 on
the spinor $(P_3,P_2)$ that we shall denote by $M_{21}$.
We write
\begin{equation}
M_{21}=
\left(\tau_{32}^{-1}\right)^{\mu_2\mu_3(\mu_1)}
\tau_{31}^{\mu_1\mu_2}\tau_{31}^{\mu_2\mu_1}
\tau_{32}^{\mu_3\mu_2(\mu_1)}.
\end{equation}
The result of the matrix multiplication
may be written in the same form as Eq.~(\ref{M1}),
except for the values of the $U(1)$ phase and of the $SU(2)$
angles:
\begin{equation}
M_{21}=e^{i\pi(\mu_1+\mu_2)}
\left(
\begin{matrix}
\phantom{-}a_{21} & b_{21}\\
-\bar b_{21} & \bar a_{21}
\end{matrix}
\right)
\ \ \mbox{with}\ \
\begin{cases}
a_{21}=\cos \pi(\mu_1+\mu_2)+i\sin\pi(\mu_1+\mu_2)\,\cos\alpha_{21},\\
b_{21}=-ie^{i\pi(\mu_1+\mu_3)}\sin\pi(\mu_1+\mu_2)\,\sin\alpha_{21},
\end{cases}
\label{M21}
\end{equation}
where
\begin{equation}
\cos\alpha_{21}=\frac{s_2s_3-s_1s_{123}}{s_{12}s_{13}},\
\sin\alpha_{21}=-\frac{2\sqrt{s_1s_2s_3s_{123}}}{s_{12}s_{13}}.
\end{equation}
The monodromy
$M_{21}$ is nothing but the exchange 
of particles \#2 and \#1 repeated twice, which
then has the following matrix representation:
\begin{equation}
\tau_{21}^{\mu_2\mu_1}\equiv \left(\tau_{32}^{-1}\right)^{\mu_3\mu_2(\mu_1)}
\tau_{31}^{\mu_2\mu_1}
\tau_{32}^{\mu_3\mu_2(\mu_1)},
\end{equation}
and whose explicit expression reads
\begin{equation}
\tau_{21}^{\mu_2\mu_1}=\frac{e^{i\pi\mu_2}}{\sqrt{s_{13}s_{23}}}
\left(
\begin{matrix}
e^{-i\pi(\mu_1+\mu_3)}\sqrt{s_1s_2} & - e^{i\pi(\mu_1+\mu_3)}\sqrt{s_3s_{123}}\\
-e^{-i\pi(\mu_1+\mu_3)}\sqrt{s_3s_{123}} & - e^{i\pi(\mu_1+\mu_3)}\sqrt{s_1s_{2}}
\end{matrix}\right).
\label{eq:P21tilde}
\end{equation}

\subsubsection{Braids}

There are interesting relations between the
transformations~(\ref{M0}),(\ref{M1}),(\ref{M21}),%
(\ref{eq:P31tilde}),(\ref{eq:P32tilde}),(\ref{eq:P21tilde})
which seem to point to some 
underlying braid group
structure \cite{A}.
(It was anticipated by 't Hooft in Ref.~\cite{H} that
braids and/or knots would play a r\^ole in this problem).
We mention here the correspondence between the monodromy
group and the braid group since it could be important
for quantization, but without entering the details:
We shall refer the interested
reader to the litterature for
an introduction to braids.

Let us consider 3 strings,
each of them being
attached to one of the particles, and which carry
the mass of the latter.
We assign the relabeling matrices $\tau$ to the crossings
of two out of the three 
strands.
For each crossing,
there are two possibilities, depending on which
one of the
particles passes above.
We choose the convention that the matrix
$\tau_{32}^{\mu_i\mu_j(\mu_k)}$
moves the strand from 
the middle of the braid 
associated to particle \#i of mass $\mu_i$
above the strand initially 
at the bottom
associated to particle \#j. The matrix
 $\tau_{31}^{\mu_i\mu_k}$ 
takes the strand at the top
above the strand initially in the middle.
The opposite orderings
are represented by the inverses of the $\tau$ matrices, namely
\begin{equation}
(\tau_{32}^{-1})^{\mu_i\mu_j(\mu_k)}\equiv
(\tau_{32}^{\mu_j\mu_i(\mu_k)})^\dagger\ \
\mbox{and}\ \
(\tau_{31}^{-1})^{\mu_i\mu_k}\equiv
(\tau_{31}^{\mu_k\mu_i})^\dagger.
\end{equation}
Note that one has to keep proper track of the particle
masses along the
strands.

Now we check by an explicit calculation that
the following relation hold:
\begin{equation}
\tau_{31}^{\mu_3\mu_2}\tau_{32}^{\mu_1\mu_2(\mu_3)}\tau_{31}^{\mu_3\mu_1}
=\frac{e^{i\pi\mu}}{\sqrt{s_{13}s_{23}}}
\left(\begin{matrix}
\phantom{-}\sqrt{s_1s_2}&-\sqrt{s_3s_{123}}\\
-\sqrt{s_3s_{123}}&-\sqrt{s_1s_2}
\end{matrix}\right)
=\tau_{32}^{\mu_1\mu_3(\mu_2)}\tau_{31}^{\mu_2\mu_1}\tau_{32}^{\mu_3\mu_2(\mu_1)}.
\end{equation}
This identity is reminiscent of the defining property
of the $B_3$ group of braids on three strands, see Ref.~\cite{A}.

Given the relation between monodromy
and relabeling transformations,
it is not difficult to check that
the group generated multiplicatively
by the matrices $M_{31}$, $M_{32}$ and $M_{21}$
is homomorphic to the pure braid group on three strings.
The latter is the subgroup of the braid group $B_3$ which
preserves the ordering of the strands.
We check by explicit matrix multiplication
that the two following defining relations 
of the pure braid group (see the corresponding
equation in Ref.~\cite{A}, which differs only by
an appropriate relabeling)
are identically verified:
\begin{equation}
M_{31}M_{32}M_{31}^{-1}=M_{21}^{-1}M_{32}M_{21}\ ,\ \
M_{31}M_{21}M_{31}^{-1}=M_{21}^{-1}M_{32}^{-1}M_{21}M_{32}M_{21}.
\end{equation}

\section{\label{sec5}Equations of motion and conservation laws}

So far, we have derived and studied the Hamiltonian which
describes the evolution of three pointlike particles
(see Eq.~(\ref{H})). We are now going
to investigate deeper the dynamics of the system.
From the Hamilton-Jacobi equations, we will be able to
compute the time evolution of the dilation factor
(i.e. also of the total angular momentum)
(Sec.~\ref{sec:HJ}). We will show that the momenta
$P_3$ and $P_2$ in terms of which the Hamiltonian is written
are conserved (Sec.~\ref{sec:P}), and we will define the
positions $Z_3$ and $Z_2$ 
canonically conjugate to $P_3$ and $P_2$ respectively
(Sec.~\ref{sec:Z}).

In order to avoid to have to carry along
$2\kappa^2$ factors,
let us take the momenta dimensionless
by setting $2\kappa^2\equiv 1$.

\subsection{\label{sec:HJ}Hamilton-Jacobi equations 
and time evolution of the dilation factor}

In our definition of complex positions and momenta (see Sec.~\ref{sec2}),
the Hamilton-Jacobi equations for the time evolution of the coordinates read
\begin{equation}
\dot z_2=\frac{\partial H}{\partial p_2},\ \
\dot z_3=\frac{\partial H}{\partial p_3},
\end{equation}
that is, from Eqs.~(\ref{H}) and~(\ref{P}):
\begin{equation}
\dot z_2=z_2^\mu \frac{\alpha \bar P_2+\gamma \bar P_3}
{|P_2|^2+|P_3|^2},\ \
\dot z_3=z_2^\mu \frac{\beta \bar P_2+\delta \bar P_3}
{|P_2|^2+|P_3|^2}.
\end{equation}
Comparing to the definitions~(\ref{P}), one immediately sees that
\begin{equation}
p_2\dot z_2+p_3\dot z_3=1.
\label{D1}
\end{equation}

Similarly, 
the Hamilton-Jacobi equations for the evolution of the momenta read
\begin{equation}
\dot p_2=-\frac{\partial H}{\partial z_2},\ \
\dot p_3=-\frac{\partial H}{\partial z_3}.
\end{equation}
Replacing $H$ by its expression~(\ref{H}), 
one gets, after the further
replacements of
$P_2$ and $P_3$ by Eq.~(\ref{P}):
\begin{equation}
\begin{split}
\dot p_2&=
\frac{\mu}{z_2}
-\frac{z_2^\mu}{|P_2|^2+|P_3|^2}\left(\frac{\partial F_{2}}{\partial z_2}\bar P_2
+\frac{\partial F_{3}}{\partial z_2}\bar P_3\right)\\
\dot p_3&=
-\frac{z_2^\mu}{|P_2|^2+|P_3|^2}\left(
\frac{\partial F_{2}}{\partial z_3}\bar P_2
+\frac{\partial F_{3}}{\partial z_3}\bar P_3
\right).
\end{split}
\end{equation}
Since $F_{2}$ and $F_{3}$ only depend on $\zeta$, 
their derivatives with respect to $z_2$ and $z_3$
read
\begin{equation}
\frac{\partial}{\partial
  z_2}=-\frac{\zeta}{z_2}\frac{\partial}{\partial\zeta},\ \
\frac{\partial}{\partial
  z_3}=\frac{1}{z_2}\frac{\partial}{\partial\zeta}.
\end{equation}
We then see that the following conservation law is satisfied:
\begin{equation}
\dot p_2 z_2+\dot p_3 z_3=-\mu.
\label{D2}
\end{equation}
Combining Eq.~(\ref{D1}) and~(\ref{D2}), we find that
the dilation factor $D$ defined in Eq.~(\ref{D})
has a linear evolution with the time $t$:
\begin{equation}
\dot D=1-\mu.
\end{equation}

\subsection{\label{sec:P}Cartesian momenta}

We are going to prove that $P_2$ and $P_3$ are
constants of motion.
To this aim, we need to compute
the Poisson brackets of the Hamiltonian with
the momenta:
\begin{equation}
\dot P_2=-\{H,P_2\}=\sum_{i=2,3}\left(
-\frac{\partial H}{\partial z_i}
\frac{\partial P_2}{\partial p_i}
+\frac{\partial H}{\partial p_i}
\frac{\partial P_2}{\partial z_i}
\right)
\end{equation}
and similarly $\dot P_3=-\{H,P_3\}$. We easily
find
\begin{equation}
\dot P_2=\frac{\{P_2,P_3\}\bar P_3}{|P_2|^2+|P_3|^2},\ \
\dot P_3=\frac{\{P_3,P_2\}\bar P_2}{|P_2|^2+|P_3|^2}.
\label{Pdot}
\end{equation}
We are going to show that the Poisson bracket $\{P_2,P_3\}$ 
vanishes. Replacing $P_2$ and $P_3$ by their 
expressions~(\ref{P}), we get
\begin{multline}
\{P_2,P_3\}=z_2^{2\mu-1}
\bigg\{
\left[\alpha ^\prime (\delta-\gamma\zeta)-\gamma
^\prime(\beta-\alpha\zeta)\right]p_2\\
+\left[\beta ^\prime (\delta-\gamma\zeta)-\delta
^\prime(\beta-\alpha\zeta)
-\mu(\alpha\delta-\beta\gamma)\right]p_3
\bigg\}.
\label{P2P3}
\end{multline}
From the definitions of $\alpha$, $\beta$, $\gamma$, $\delta$
in Eq.~(\ref{greeks}),
the following identities hold:
\begin{equation}
\begin{split}
\beta-\alpha\zeta&=-\sqrt{N_2}\zeta(1-\zeta)f_{2a}\\
\delta-\gamma\zeta&=\phantom{-}\sqrt{N_3}\zeta(1-\zeta)f_{3a}\\
\alpha\delta-\beta\gamma&=\phantom{-}\sqrt{N_2 N_3}\zeta(1-\zeta)
(f_{2a}f_{3b}+f_{2b}f_{3a}).
\end{split}
\label{idgreeks}
\end{equation}
The derivatives of $\alpha$, $\beta$, $\gamma$, $\delta$ can be
expressed as a function of the $f$'s and the second derivatives of
$f_{2b}$ and $f_{3b}$ by 
using
\begin{equation}
f^\prime_{2b}=-\mu_3 f_{2a},\ \
f^\prime_{3b}=\mu_3 f_{3a}
\label{derivf}
\end{equation}
which are consequences of 
standard identities between hypergeometric
functions.

We insert
Eqs.~(\ref{idgreeks}),(\ref{derivf}) into  Eq.~(\ref{P2P3})
and get
\begin{multline}
\{P_2,P_3\}=\frac{z_2^{2\mu-1}\sqrt{N_2N_3}\zeta(1-\zeta)}{\mu_3}
\bigg\{
f_{2a}\left[\zeta(1-\zeta)f ^{\prime\prime}_{3b}+\mu_3(1-\mu)f_{3b}\right]\\
+f_{3a}\left[\zeta(1-\zeta)f ^{\prime\prime}_{2b}+\mu_3(1-\mu)f_{2b}\right]
\bigg\}.
\end{multline}
Thanks to
the hypergeometric equation~(\ref{hyper}) applied to $f_{3b}$ and
$f_{2b}$,
we see that the term
under the square brackets cancels identically, and thus $\{P_2,P_3\}=0$.
Equations~(\ref{Pdot}) 
eventually show that  $P_2$ and $P_3$ are constants
of motion:
\begin{equation}
\dot P_2=0,\ \dot P_3=0.
\end{equation}

\subsection{\label{sec:Z}Cartesian positions}

We have just exhibited two constants of motion:
The two relative Cartesian momenta of the particles.
The Cartesian velocities, which are the
time derivatives of the Cartesian positions
$Z_2$ and $Z_3$, should also be constant.
We are going to define them as 
the variables conjugate to 
the momenta $P_2$ and $P_3$.

The dilation factor $D=p_2z_2+p_3z_3$ 
may be expressed with the help of $P_2$ and $P_3$ in the form
\begin{equation}
D=z_2^{1-\mu}
\left(\frac{\delta-\gamma\zeta}{\alpha\delta-\beta\gamma}P_2
-\frac{\beta-\alpha\zeta}{\alpha\delta-\beta\gamma}P_3\right).
\label{eq:DP}
\end{equation}
We define the new variables $Z_2$ and $Z_3$ in such a way that
\begin{equation}
D=(1-\mu)(Z_2 P_2+Z_3 P_3).
\label{eq:DdefZ}
\end{equation}
Let us work out the explicit expression for $Z_2$ and
$Z_3$ as a function of the canonical variables.
We introduce the notation
\begin{equation}
{{\cal W}}=\sqrt{N_2 N_3}(f_{2a} f_{3b}+f_{2b} f_{3a}),
\label{wronskian}
\end{equation}
which is the inner product of the spinors 
$\sigma_a$ and
$\sigma_b$, namely
\begin{equation}
{\cal W}=(\sigma_b,\sigma_a)\equiv\sigma_b^T
\left(\begin{matrix}
0 & -1\\1 & 0
\end{matrix}\right)
\sigma_a
\end{equation}
and thus a $SU(2)$ invariant.
It turns out that ${\cal W}$ has a simple expression, 
see Eq.~(5.5) in Ref.~\cite{BCV2}: 
As a matter of fact, it is a Wronskian function for
the solutions of a second-order differential equation.

After comparison of Eqs.~(\ref{eq:DP}) and~(\ref{eq:DdefZ}),
the explicit expressions for $Z_2$ and $Z_3$ are
\begin{equation}
\begin{split}
Z_2&=z_2^{1-\mu}\frac{\sqrt{N_3} f_{3a}}{(1-\mu){{\cal W}}}\\
Z_3&=z_2^{1-\mu}\frac{\sqrt{N_2} f_{2a}}{(1-\mu){{\cal W}}}.
\end{split}
\label{Z}
\end{equation}
The pair $Z\equiv (Z_2,-Z_3)$ makes up a spinor, 
in such a way that the spinorial product
$(P,Z)=P_2 Z_2+P_3 Z_3$ be an invariant under $SU(2)$ transformations.

In order to make contact with our intuition of classical physics,
let us again take the small mass limit in which gravitational effects
vanish.
Then a straightforward calculation leads to
\begin{equation}
Z_3=\sqrt{N_3}\,z_3,\ Z_2=\sqrt{N_2}\left(z_2-\frac{\mu_3}{\mu_{13}}z_3\right),
\label{eq:Zsmallm}
\end{equation}
that is to say, $Z_3$ is the position of particle \#3
up to a normalization,
and $Z_2$ the one of particle \#2 with respect to the system of particles
$(13)$. (We recall that we have chosen a frame in which $z_1=0$).

The dilation factor
$D$ has a constant time derivative and the $P_n$ are constants of
motion,
we expect that the $Z_n$ 
also have constant time derivatives.
This turns out to be true, and furthermore, the $Z_n$' s 
are canonically conjugate to $P_n$'s.
The first point can be shown by evaluating
\begin{equation}
\dot Z_n=-\{H,Z_n\}=\frac{\{Z_n,P_2\}\bar P_2
+\{Z_n,P_3\}\bar P_3 }{|P_2|^2+|P_3|^2}.
\label{eq:Zndot}
\end{equation}
Let us perform the calculation completely for $n=2$.
The Poisson bracket of $Z_2$ and $P_2$ that appears 
in Eq.~(\ref{eq:Zndot})
reads
\begin{equation}
\{Z_2,P_2\}=z_2^{\mu-1}\left[(1-\mu)\alpha
Z_2+(\beta-\alpha\zeta){Z_2^\prime}\right].
\label{ZP}
\end{equation}
To compute the derivative of $Z$, we use the more general formula
\begin{equation}
\frac{d}{d\zeta}\left(\frac{f^\prime}{{\cal W}}\right)
=-\frac{\mu_3(1-\mu)}{\zeta(1-\zeta)}\frac{f}{{{\cal W}}},
\label{theorem}
\end{equation}
valid for any solution $f$ of the hypergeometric
equation~(\ref{hyper}).
We introduce the derivatives of $f_{3b}$ in Eq.~(\ref{Z})
expressed in 
Eq.~(\ref{derivf}). One then uses the previous formula
with $f$ set to $f_{3b}$
to compute $Z_2^\prime$ in Eq.~(\ref{ZP}). With the help of the 
identities~(\ref{idgreeks}),
one arrives at
\begin{equation}
\{Z_2,P_2\}=1.
\label{eq:Poisson1}
\end{equation}
Similar calculations lead to the following Poisson brackets:
\begin{equation}
\{Z_2,P_3\}=0,\ \{Z_3,P_2\}=0,\ \{Z_3,P_3\}=1,\ \{Z_2,Z_3\}=0,\ 
\{P_2,P_3\}=0.
\label{eq:Poisson2}
\end{equation}
Thus,
\begin{equation}
\dot Z_2=\frac{\bar P_2}{|P_2|^2+|P_3|^2},\ 
\dot Z_3=\frac{\bar P_3}{|P_2|^2+|P_3|^2}.
\end{equation}
For consistency, we easily check that $\dot D=1-\mu$.

It is possible to recover the usual
Cartesian equations of motion
\begin{equation}
\frac{dZ_2}{dT}=2\bar P_2,\  \frac{dZ_3}{dT}=2\bar P_3
\label{eq:Cartesianmotion}
\end{equation}
by changing the time gauge as follows:
\begin{equation}
\frac{dt}{dT}=2(|P_2|^2+|P_3|^2).
\end{equation}
This reparametrization depends on 
the variables $z_n$ and $p_n$.
With this choice for time and taking
$Z_n$ and $P_n$ as canonical variables, 
the $T$-evolution would be
given by the Hamiltonian
\begin{equation}
E=2(|P_2|^2+|P_3|^2),
\end{equation}
up to a constant.
Such a Hamiltonian would be simpler and
more intuitive since it is the nonrelativistic
kinetic energy of the three particles in the center-of-mass frame
(see Eq.~(\ref{eq:E}) and the discussion above it), 
but the phase-space variables would not be single valued.

\section{\label{sec6}
Consistency with previous
calculations and tentative extension to many bodies}

In the present work,
the computation of the Hamiltonian is based on
a calculation of the total mass 
$\mu=\sum_n\mu_n+\varepsilon$ as a function of
the masses, positions and momenta of the particles in 
the framework of the second-order formalism.

On the other hand, since (2+1) gravity is a topological theory,
we know that
pointlike particles move on straight lines
with constant velocities $V_n$, when appropriate (Cartesian) 
coordinates are chosen.
The total mass $\mu$ can be obtained by
writing the total Cartesian momentum of the system of the 3 particles.
The result for $\mu$ should be the same as the one obtained
in the previous section: This is what we are going to
check here (Sec.~\ref{sec:consistency}).
This computation will also help us to establish the
relationship between
$V_n$ and the derivatives of $Z_n$.
It can be extended to four (or more) particles
and allow us to guess the form of the Hamiltonian
in these cases (Sec.~\ref{sec:extension}).


\subsection{\label{sec:consistency}
Comparison with the first-order formalism}

Following Ref.~\cite{BCV2}, we introduce a spin-$\frac12$
representation of the holonomy related to transport
on a curve around the particle $n$ of mass $\mu_n$
and velocity $V_n$ in the form
\begin{equation}
L_n(\mu_n,V_n)=
\left(
\begin{matrix}
\bar a_n & b_n\\
\bar b_n & a_n
\end{matrix}
\right)\ ,
\ \ \mbox{where}\ \
\begin{cases}
a_n=\cos\pi\mu_n+i\gamma_n\sin\pi\mu_n,\\
b_n=-i\gamma_n \bar V_n\sin\pi\mu_n,
\end{cases}
\end{equation}
and $\gamma_n=1/{\scriptstyle \sqrt{1-|V_n|^2}}$.
In order to ``measure'' the total mass of the system,
we may travel on a loop that goes around all the three particles.
The total mass of the system $\mu$ is then computed
from the trace of the holonomy given by
the product of the three matrices $L_n$:
\begin{equation}
\cos \pi\mu=\frac12 \mbox{Tr} 
[L_3(\mu_3,V_3)L_2(\mu_2,V_2)L_1(\mu_1,V_1)].
\label{first}
\end{equation}
The order of the product is determined by
the choice of ordering the particles anticlockwise in space.
The calculation leads to
\begin{multline}
\cos\pi\mu=\cos\pi\mu_1\cos\pi\mu_2\cos\pi\mu_3\\
-\left(\gamma_1\gamma_2(1-\vec V_1\cdot\vec V_2)
\sin\pi\mu_1\sin\pi\mu_2\cos\pi\mu_3+
\mbox{[cyclic permutations]}\right)\\
+\frac{i}{2}\gamma_1\gamma_2\gamma_3
(-V_{23}\bar V_1+V_{13}\bar V_2-V_{12}\bar V_3)
\sin\pi\mu_1\sin\pi\mu_2\sin\pi\mu_3.
\end{multline}
We now expand to the lowest order in the velocities.
We observe that the result can be written as
the sum of the masses of the particles
and of a quadratic form of the velocities,
representing the total nonrelativistic kinetic energy of the system:
\begin{equation}
\mu=\mu_1+\mu_2+\mu_3+\frac{1}{\pi}
(\bar V_{12}\ \bar V_{13})
Q\left(
\begin{matrix}V_{12}\\V_{13}\end{matrix}
\right),
\end{equation}
where
\begin{equation}
Q=\frac{1}{2s_{123}}\left(
\begin{matrix}
 s_{2}s_{13} & -e^{-i\pi\mu_1}s_2 s_3\\
-e^{i\pi\mu_1}s_2 s_3&{s_{12}s_3}
\end{matrix}
\right).
\end{equation}
We may perform a standard
decomposition of $Q$ in a product of
lower diagonal $L$, diagonal $D$ and upper diagonal $U$
matrices $Q=LDU$,
with in this case,
\begin{equation}
D=\frac12\left(
\begin{matrix}
\frac{s_2 s_{13}}{s_{123}}&0\\
0&\frac{s_1 s_{3}}{s_{13}}
\end{matrix}
\right),\
U=\left(
\begin{matrix}
1 & -e^{-i\pi\mu_1}\frac{s_3}{s_{13}}\\
0 & 1
\end{matrix}
\right)\ \mbox{and}\
L=U^\dagger
\end{equation}
since $Q$ is Hermitian.
From this decomposition, we immediately 
write the kinetic energy as a sum of squared moduli:
\begin{equation}
\mu-\sum_n \mu_n=\frac{1}{2}\bigg(
\frac{s_2 s_{13}}{\pi s_{123}}\left|
\bar V_{12}-\frac{s_3 e^{i\pi\mu_1}}{s_{13}}
\bar V_{13}
\right|^2+\frac{s_1s_3}{\pi s_{13}}|V_{13}|^2
\bigg).
\label{lorentzfinal}
\end{equation}
Note that the procedure used to arrive at this
factorized expression may be repeated for a number of particles
larger than 3. We sketch it in the
next subsection~\ref{sec:extension}.

A result of the first-order formalism developed
in Ref.~\cite{BCV2} is that
the Cartesian velocities $V_n$ can then be expressed
with the help of the canonical positions $z_n$ of the particles
(see Eq.~(4.19) in Ref.~\cite{BCV2}). They read
\begin{equation}
\bar V_{1n}=2 K_0 z_2^{\mu-1}
\int_0^{\zeta_n}d\xi\,\xi^{\mu_1-1}(\xi-1)^{\mu_2-1}(\xi-\zeta)^{\mu_3-1}
(\xi-\eta_A),
\label{firstbis0}
\end{equation}
where $\eta_A$ is the apparent singularity.
From the integral representation of the
hypergeometric function
\begin{equation}
\int_0^1  dz z^{\alpha-1}
(1-z)^{\beta-1}(1-tz)^{\gamma-1}\equiv B_{\alpha,\beta}
F(\alpha,1-\gamma,\alpha+\beta,t),
\label{eq:defhyper0}
\end{equation}
and after appropriate changes of variables,
we express the $\bar V$'s in terms of 
the $f$'s defined in Eq.~(\ref{f}), namely
\begin{equation}
\begin{split}
\bar V_{12}&=-2 K_0 z_2^{\mu-1}e^{-i\pi\mu_2}\left\{
\frac{s_3}{s_{13}}e^{i\pi(\mu_1+\mu_3)}[f_{3b}+(\eta_A-\zeta)f_{3a}]
+[f_{2b}-(\eta_A-\zeta)f_{2a}]\right\}\\
\bar V_{13}&=-2 K_0 z_2^{\mu-1}
e^{i\pi(\mu_3-\mu_2)}\left[f_{3b}+(\eta_A-\zeta)f_{3a}\right].
\label{firstbis}
\end{split}
\end{equation}
Inserting these expressions for the velocities in Eq.~(\ref{lorentzfinal}),
we find
\begin{equation}
\mu-\sum_n\mu_n={2|K_0|^2}|z_2|^{2\mu-2} \left(
 \frac{s_1s_3}{\pi s_{13}}|(\eta_A-\zeta)f_{3a}+f_{3b}|^2
+\frac{s_2s_{13}}{\pi s_{123}}|(\eta_A-\zeta)f_{2a}-f_{2b}|^2
\right)
\label{lorentzend}
\end{equation}
Thus $\varepsilon$ in Eq.~(\ref{finalepsilon})
matches $\mu-\sum\mu_i$ given by Eq.~(\ref{lorentzend})
from the first-order formalism provided 
\begin{equation}
\eta_A=\zeta_A\ \
\mbox{and}\ \ 
|K_0|^2=\frac{|K|^2}{8}.
\end{equation}
Now we can establish the relation between the velocities,
the Cartesian momenta and coordinates.
Identifying Eq.~(\ref{firstbis}) with the expressions
of the momenta~(\ref{P0}), we write
\begin{equation}
\begin{split}
\bar V_{31}&=\frac{e^{i\pi(\mu_3-\mu_2)}}{\sqrt{2}}\frac{K}{D}P_3,\\
\bar V_{21}&=\frac{e^{i\pi(\mu_3-\mu_2)}}{\sqrt{2}}\frac{K}{D}
\left(e^{-i\pi\mu_3}\frac{P_2}{\sqrt{N_2}}
+e^{i\pi\mu_1}\frac{s_3}{s_{13}}\frac{P_3}{\sqrt{N_3}}
\right).
\end{split}
\end{equation}
On the other hand, by definition of the Cartesian momenta $P_2$ and $P_3$,
\begin{equation}
\frac{|D|^2}{|K|^2}=\frac{|P_2|^2+|P_3|^2}{4\varepsilon}.
\end{equation}
The right-hand side is a numerical constant since
$\varepsilon$ has to be identified with the nonrelativistic
kinetic energy
in the Cartesian time gauge.
Therefore
\begin{equation}
\begin{split}
\bar V_{31}&={e^{i\phi}}\frac{2P_3}{\sqrt{N_3}},\\
\bar V_{21}&={e^{i\phi}}
\left(e^{-i\pi\mu_3}\frac{2P_2}{\sqrt{N_2}}
+e^{i\pi\mu_1}\frac{s_3}{s_{13}}\frac{2P_3}{\sqrt{N_3}}
\right),
\end{split}
\end{equation}
where we have introduced the angle $\phi$ to
absorb all irrelevant phases.
One may now replace $P_2$ and $P_3$ by the derivative
of $Z_2$ and $Z_3$ with respect to $T$ 
(see Eq.~(\ref{eq:Cartesianmotion})).
Taking the small mass limit, with the help Eq.~(\ref{eq:Zsmallm}), 
it is easy to see that $\bar V_{31}$ and $\bar V_{21}$
coincide in this limit with
the derivatives of $z_3$ and $z_2$ respectively.


\subsection{\label{sec:extension}
Guessing the form of the Hamiltonian for many bodies}

As was suggested before, we may express
the total kinetic energy $\varepsilon$
of a system of many bodies as a function
of the Cartesian momenta of specific 
subsystems of particles.
Then from a change of the time gauge, we may infer
the form of the Hamiltonian.
We address in some
detail the case of the four-body system.
We write a relation similar to Eq.~(\ref{first}):
\begin{equation}
\cos \pi\mu=\frac12 \mbox{Tr} [L_4(\mu_4,V_4)
L_3(\mu_3,V_3)L_2(\mu_2,V_2)L_1(\mu_1,V_1)],
\label{first4bodies}
\end{equation}
which we then expand at lowest order in the Cartesian velocities $V_{1i}$
\begin{equation}
\mu=\mu_1+\mu_2+\mu_3+\mu_4+\frac{1}{\pi}
(\bar V_{12}\ \bar V_{13}\ \bar V_{14})
Q\left(
\begin{matrix}V_{12}\\V_{13}\\V_{14}\end{matrix}
\right).
\end{equation}
In this case, the matrix $Q$ encoding the
quadratic form of the velocities reads
\begin{equation}
Q=\frac{1}{2s_{1234}}\left(\begin{matrix}
s_2s_{134} & -e^{-i\pi\mu_{14}}s_2s_3 &-e^{-i\pi(\mu_1-\mu_3)}s_2s_4\\
-e^{i\pi\mu_{14}}s_2s_3 & s_3s_{124}  & -e^{-i\pi\mu_{12}}s_3s_4\\
-e^{i\pi(\mu_1-\mu_3)}s_2s_4 & -e^{i\pi\mu_{12}}s_3s_4 & s_4s_{123}
\end{matrix}\right).
\end{equation}
Performing as in the three-body case
a $Q=LDU$ decomposition
we find
the following formula
for $\varepsilon\equiv\mu-\sum_n\mu_n$:
\begin{multline}
\varepsilon=\frac{1}{2}\left(
\frac{s_2s_{134}}{\pi s_{1234}}
\left|\bar V_{12}-e^{i\pi(\mu_1+\mu_4)}\frac{s_3}{s_{134}}\bar V_{13}
-e^{i\pi(\mu_1-\mu_3)}\frac{s_4}{s_{134}}\bar V_{14}\right|^2\right.\\
\left.+\frac{s_3s_{14}}{\pi s_{134}}
\left|\bar V_{13}-e^{i\pi\mu_1}\frac{s_4}{s_{14}}\bar V_{14}
\right|^2
+\frac{s_1s_{4}}{\pi s_{14}}
\left|\bar V_{14}
\right|^2
\right).
\end{multline}
The identification with the classical equation for the total kinetic energy
of the system (in the Cartesian time gauge)
\begin{equation}
\varepsilon=
2\left(
|P_2|^2+|P_3|^2+|P_4|^2
\right)
\end{equation}
leads (up to phases)
to an expression of the Cartesian momenta $P_4$, $P_3$,
$P_2$ of particles number $\#4$ relatively to \#1, $\#3$ relatively
to the system $(14)$, and $\#2$ relatively to the system $(134)$
respectively.
This statement can be checked
in the small mass limit in which the phases
become irrelevant.
We find for $P_4$, $P_3$~and~$P_2$
formulas similar to~(\ref{Plimit})
and which, up to normalization factors,
are the expressions of the momenta of free 
systems of particles as a function
of their Cartesian velocities.

The strong similarity with the $3$-body problem suggests that
the Hamiltonian for $N=4$ particles
and more generally, for
an arbitrary number $N$ of particles,
would read
\begin{equation}
H=\ln\left(
\frac{|P_2|^2+|P_3|^2+\cdots+|P_N|^2}{4\varepsilon}
\right).
\end{equation}
Note however
that the expressions of the $P_n$ (or $V_{1n}$) in the regular phase
space coordinates
$(z_n,p_n)$ would involve line integrals in the complex plane with 
$N$ cuts,
which are not expected to be related in a simple way to known functions,
as they are in the 2- and 3-body cases.

\section{\label{sec7}Towards the quantum Hilbert space}

We impose canonical quantization 
by replacing the Poisson brackets $\{\cdot,\cdot\}$ by the commutators
$-i\hbar[\cdot,\cdot]$, which is realized by the substitution 
$p_{j}\rightarrow -i\hbar\partial_{z_{j}}$.
In units $\hbar=c=1$, this results in the usual relations
between the positions $z_i$ and momenta $p_j$, namely
\begin{equation}
[z_i,p_j]=i\delta_{ij},\ [z_i,z_j]=0,\ [p_i,p_j]=0.
\end{equation}
It turns out that the same rules apply to
$Z_i$ and $P_j$, provided the ordering of $z_i$ and $p_j$
is the one given in the defining equation~(\ref{P}) (i.e.
the position operators are to the left of the momentum
operators).

The problem then arises of finding an explicit
expression for the eigenfunctions of the Hamiltonian,
so as to construct the quantum Hilbert space and the scattering
amplitudes. It turns out that it is not difficult to
find explicit solutions for the Schr\"odinger equation,
because the Hamiltonian is a simple function
of the $P_2$, $P_3$ (and $\bar P_2$, $\bar P_3$) operators,
which are diagonalized by the ``plane waves''
\begin{equation}
\psi_{k_2,k_3}(z_2,z_3)=e^{i[k_2 Z_2(z_j)+k_3 Z_3(z_j)
+(k_j\leftrightarrow\bar k_j, Z_j\leftrightarrow \bar Z_j)]},
\label{eq:wavefunction}
\end{equation}
where $k_2$, $k_3$ label the eigenvalues of $P_2$, $P_3$
and $Z_2$, $Z_3$ are expressed in terms of the
regular coordinates $z_j$ by Eq.~(\ref{Z}).
The corresponding energy eigenvalue of Eq.~(\ref{eq:Hdiffkinetic}) 
is $E=2(|k_2|^2+|k_3|^2)$.

However, the wave functions~(\ref{eq:wavefunction})
are not single valued, but transform in a way induced
by the monodromy transformations $M$ of Sec.~\ref{sec:monobraids}.
In fact, if we let $z_2\rightarrow z_2^M=z_2 e^{2i\pi}$ or 
 $z_3\rightarrow z_3^M=z_3 e^{2i\pi}$ or combinations thereof,
then $(Z_2,-Z_3)$ transforms as a spinor. 
The invariant combination in the exponent of
(\ref{eq:wavefunction}) translates this transformation 
into a $U(2)$
transformation of the momenta: 
$k\equiv(k_3,k_2)\rightarrow k^M=(k_3^M,k_2^M)$.
The transformed wave function reads
\begin{equation}
\psi_{k}(z_j^M)=e^{i[k_2^M Z_2(z_j)+k_3^M Z_3(z_j)+\mbox{c.c.}]}.
\end{equation}
This wave function possesses the same energy
eigenvalue $E=2( |k_2|^2+|k_3|^2)$ 
as $\psi_{k}(z_j)$ in Eq.~(\ref{eq:wavefunction}).
In other words, $\psi_{k}(z_j)$ transforms into another
wave function of the same degeneracy class
according to a $U(2)$ representation of the monodromy.

A similar situation arises already at the level of the simple
two-body case, where the analog of Eq.~(\ref{eq:wavefunction})
is the wave function
\begin{equation}
\psi_k(z)=e^{i[kZ(z)+\mbox{c.c.}]}=
e^{i\left[k {z^{1-\mu}}/({1-\mu})+\mbox{c.c.}\right]}\ ,\ \
E=2|k|^2.
\end{equation}
Similarly, it is not monodromic, but transforms by a $U(1)$ monodromy
as
\begin{equation}
\psi_k(z)\rightarrow
\psi_k(e^{2i\pi}z)=\psi_{ke^{2i\pi\alpha}}(z)\ ,\ \ (\alpha=1-\mu),
\end{equation}
which remains in the same degeneracy class.
One may think that 
in the two-body $U(1)$ case, it would be 
possible to construct
linear combinations of plane waves
which are single valued and have the same energy by just
summing with unit weight over the whole monodromy group,
as follows:
\begin{equation}
\tilde\psi_k(z)=\sum_{n=-\infty}^{+\infty}\psi_k(e^{2i\pi n}z).
\label{eq:sumovern}
\end{equation}
However, this turns out to be too naive.
Unless the monodromy group is finite (for fractional values
of $\alpha$), the series sums up to zero, after
appropriate regularization.

A solution which is both finite and monodromic was proposed
by Deser, Jackiw \cite{DJ} and 't Hooft \cite{H}.
It also consists in writing a superposition
of plane waves degenerate in energy (which
differ by the azimuthal angle of their momenta), but with
nontrivial
weights. Let us introduce some
useful notations:
\begin{equation}
x\equiv |k||z|^\alpha/\alpha\ ,\ \ 
\beta\equiv\arg \left(k\right)\ ,\
\ \theta\equiv-\arg z.
\end{equation}
Then
\begin{equation}
\psi_k(z)=\int_{{\cal C}_{\alpha\theta}}\frac{d\beta}{2\pi}
\hat f(\beta) e^{i x\cos(\beta-\alpha\theta)}
\end{equation}
solves the Schr\"odinger equation when the contour
${\cal C}_{\alpha\theta}$ is a combination of Schl\"afli
contours in the upper/lower complex plane of the $\beta$ variable,
namely the union of the broken line
\begin{equation}
{\cal C}_+\equiv]-\pi+i\infty,-\pi,\pi,\pi+i\infty[
\end{equation}
translated along the real axis by $\alpha\theta$
and of its complex conjugate ${\cal C}_-$ 
(we introduce the notation ${\cal C}\equiv
{\cal C}_+ +{\cal C}_-$ for the union), 
with appropriate deformations
in order to avoid the possible singularities of $\hat f$.
The function $\hat f$ is meromorphic and well behaved for 
$\mbox{Im}(\beta)\rightarrow\pm\infty$.
The convergence lines in $\beta$ are 
$\beta-\alpha\theta\rightarrow
(2n+1)\pi\pm i\infty$, thus
the contour ${\cal C}_{\alpha\theta}$ may be translated by
$\beta\rightarrow\beta\pm 2\pi n$.

For each choice of ${\cal C}$ one may now construct a monodromic solution by replacing $\theta$
by $\theta_n\equiv \theta+2n\pi$ and summing over $n$:
\begin{equation}
\tilde\psi_k(z)=\sum_n\int_{{\cal C}_{\alpha\theta_n}}\frac{d\beta}{2\pi}
\hat f(\beta) e^{i x \cos(\beta-\alpha\theta_n)}
=\int_{{\cal C}}\frac{d\beta}{2\pi}
\left[\sum_n \hat f(\beta+\alpha\theta_n)\right] e^{i x \cos\beta}.
\end{equation}
The function 
\begin{equation}
f(\beta)\equiv\sum_n \hat f(\beta+2\pi\alpha n)
\end{equation}
which appears under the square brackets is periodic of period
$2\pi\alpha$, so that $\psi_k$ is monodromic under the rotation
$\theta\rightarrow\theta+2\pi$.

Such a monodromic expression can then be projected on the 
integer angular
momentum $m$ as follows:
\begin{equation}
\psi_k^m(z)=
\int_{-\infty}^{+\infty}\frac{d\theta}{2\pi}e^{-im\theta}\int_{\cal C}
\frac{d\beta}{2\pi}\hat f(\beta+\alpha\theta)e^{i x \cos\beta}
=\int_{\cal C}\frac{d\beta}{2\pi}\left[\int_{-\pi}^{+\pi}\frac{d\theta}{2\pi}
f(\beta+\alpha\theta)e^{-i m\theta}\right]e^{i x \cos\beta}.
\label{eq:wavefunctionm}
\end{equation}
The integral over $\theta$ (under the square brackets) yields
the Fourier coefficient $f_m$ of the periodic function $f$
times the phase $e^{i m\beta/\alpha}$.
The integral over $\beta$ is then seen to yield 
the Bessel function
$J_{|m|/\alpha}(x)$, which is the DJH \cite{DJ,H} result. The r\^ole of the weight factor $\hat f$ is to suppress the contributions of the secondary sheets so as to avoid too much destructive interference. For the DJH scattering process
$\hat f \sim 1/(\beta-\pi\alpha)$ and $f \sim$ tan $(\beta/2\alpha)$.

Is a similar procedure available for $U(2)$ (or more generally $U(N-1)$)?
On one hand, when the monodromy group is finite,
we may write
\begin{equation}
\tilde\psi_k(z_j)=\sum_{\mbox{\footnotesize monodromies } M}\psi_{k^M}(Z(z_j))
=\sum_{M}
e^{i[k_2^M Z_2(z_j)+k_3^M Z_3(z_j)+\mbox{c.c.}]},
\label{eq:mono}
\end{equation}
where $k$  is now a $U(2)$ spinor with components $(k_3,k_2)$, 
and $Z$ has components $(Z_2,-Z_3)$.
But on the other hand, despite various attempts,
we have been unable to extend to this non-abelian case the
general harmonic analysis of Eq.~(\ref{eq:wavefunctionm}).
The main obstacle is the nonabelian group multiplicative structure
which does not allow an explict evaluation (with identification
of good quantum numbers).

Let us however give an example that shows how a
simple structure may arise for proper
(rational) values of the masses.
Let us consider the case in which the particle masses
are $\mu_1=\mu_2=\mu_3=\frac14$.
Then the three basic
monodromy matrices are related to the Pauli matrices $\sigma_1$, 
$\sigma_2$ and $\sigma_3$ through
\begin{equation}
M_{31}=-\sigma_3,\ M_{32}=\sigma_1,\ M_{21}=\sigma_2.
\end{equation}
The group generated by the monodromies is finite in this case.
It is made of the sixteen $U(2)$ matrices
$\epsilon I$, $\epsilon \sigma_1$, $\epsilon \sigma_2$, $\epsilon\sigma_3$,
where $\epsilon\in\{1,i,-1,-i\}$.

In this particular case, the Cartesian coordinates have a simple
expression as a function of the regular coordinates $z_2$ and $\zeta=z_3/z_2$.
Equations~(\ref{Z}) indeed boil down to
\begin{equation}
Z_2=\frac{4\pi}{\Gamma^2(\frac14)} z_2^{1/4}
\left(1+\sqrt{1-\zeta}\right)^{1/2}\ ,\ \ 
Z_3=\frac{4\pi}{\Gamma^2(\frac14)} z_2^{1/4}
\left(1-\sqrt{1-\zeta}\right)^{1/2}.
\end{equation}
We now apply Eq.~(\ref{eq:mono})
to obtain the explicit monodromic wave function:
\begin{multline}
\tilde\psi_k(z_2,\zeta)=\sum_{\epsilon\in\{1,i,-1,-i\}}
\bigg\{e^{i[\epsilon k_2 z_2^{1/4}\left(1+\sqrt{1-\zeta}\right)^{1/2}
+\epsilon k_3 z_2^{1/4}\left(1-\sqrt{1-\zeta}\right)^{1/2}+\mbox{c.c.}]}\\
+e^{i[\epsilon k_2 z_2^{1/4}\left(1-\sqrt{1-\zeta}\right)^{1/2}
+\epsilon k_3 z_2^{1/4}\left(1+\sqrt{1-\zeta}\right)^{1/2}+\mbox{c.c.}]}\\
+e^{i[\epsilon k_2 z_2^{1/4}\left(1+\sqrt{1-\zeta}\right)^{1/2}
-\epsilon k_3 z_2^{1/4}\left(1-\sqrt{1-\zeta}\right)^{1/2}+\mbox{c.c.}]}\\
+e^{i[-i\epsilon k_2 z_2^{1/4}\left(1-\sqrt{1-\zeta}\right)^{1/2}
+i\epsilon k_3 z_2^{1/4}\left(1+\sqrt{1-\zeta}\right)^{1/2}+\mbox{c.c.}]}
\bigg\}.
\end{multline}

\section{Summary and suggestions}

In this paper, we have fully understood the canonical Hamiltonian
structure of the 3-body problem in the nonrelativistic limit
in which the particle velocities are small.
Let us summarize our findings, repeating the explicit formulas
to which we have arrived.

We have provided the expression of the Hamiltonian in the form
\begin{equation}
H=\frac{1}{2\kappa^2}
\ln\frac{(2\kappa^2)^2\left(|P_2|^2+|P_3|^2\right)}
{4\varepsilon}
\end{equation}
where $P_2$ and $P_3$ are properly-defined relative
momenta, which are given as explicit functions
of the canonical coordinates $(z_2,p_2)$ and $(z_3,p_3)$,
while
$\varepsilon$ is a fixed parameter which does not play a r\^ole
in the dynamics:
It represents the difference between the total
(adimensionalized) mass of the Universe $\mu$,
and the sum of the masses $\mu_n$ of the particles,
which are all constants.
The nonrelativistic limit implies that it is small compared
to $\mu$.

By putting Eqs.~(\ref{P}) and~(\ref{greeks}) together, we obtain the expressions
\begin{equation}
\begin{split}
P_3&=z_2^\mu\sqrt{N_3}
\left[
f_{3b}(p_2+\zeta p_3)+f_{3a}\zeta(1-\zeta)p_3\right],\\
P_2&=z_2^\mu \sqrt{N_2}
\left[
f_{2b}(p_2+\zeta p_3)-f_{2a}\zeta(1-\zeta)p_3
\right],
\end{split}
\end{equation}
where the normalization factors $N_3$ and $N_2$ are given
by Eq.~(\ref{N}), namely
\begin{equation}
N_3=\frac{\sin\pi\mu_1 \sin\pi\mu_{3}}{\pi\sin\pi\mu_{13}},\
N_2=\frac{\sin\pi\mu_2 \sin\pi\mu_{13}}{\pi\sin\pi\mu},
\end{equation}
where $\mu_{ij}=\mu_i+\mu_j$ and $\mu=\mu_1+\mu_2+\mu_3$.
The $f$'s are solutions of the hypergeometric
equation with specific coefficients (see Eq.~(\ref{hyper})),
and their expressions read~(\ref{f})
\begin{equation}
\begin{split}
f_{3a}&=\frac{\Gamma(\mu_1)\Gamma(\mu_3)}{\Gamma(\mu_{13})}
\zeta^{\mu_{13}-1}
F(\mu_1,1-\mu_2,\mu_{13},\zeta),\\
f_{2a}&=\frac{1-\mu}{1-\mu_{13}}
\frac{\Gamma(\mu_{13})\Gamma(\mu_2)}{\Gamma(\mu)}
F(1-\mu_3,2-\mu,2-\mu_{13},\zeta),
\end{split}
\label{fbis}
\end{equation}
and $f_{3b}$, $f_{2b}$ are obtained from $f_{3a}$, $f_{2a}$ respectively
by shifting $\mu_3$ to $\mu_3+1$.
$P_3$ and $P_2$ are constants of motion which are
related to the Cartesian momenta.

One can also build the Cartesian coordinates, which have
relatively simple expressions. From Eq.~(\ref{Z}),
we get
\begin{equation}
\begin{split}
Z_2&=\frac{z_2^{1-\mu}}{1-\mu}
\frac{\Gamma(\mu)}{\sqrt{N_2}\Gamma(\mu_{13})\Gamma(\mu_2)}
F(\mu_3,\mu-1,\mu_{13},\zeta),\\
Z_3&=\frac{z_2^{1-\mu}}{1-\mu_{13}}
\frac{\Gamma(\mu_{13})}{\sqrt{N_3}\Gamma(\mu_1)\Gamma(\mu_3)}
\zeta^{1-\mu_{13}}
F(\mu_2,1-\mu_1,2-\mu_{13},\zeta).
\end{split}
\end{equation}
Finally, the Cartesian time $T$ is related to the ADM time $t$
by
\begin{equation}
\frac{dT}{dt}=\frac{1}{2(2\kappa^2)^2(|P_2|^2+|P_3|^2)},
\end{equation}
which represents a time-gauge change.

An important r\^ole is played by the $U(2)$ symmetry of $H$,
under which $(P_3,P_2)$ and $(Z_2,-Z_3)$ transform as spinors.
This symmetry regulates the monodromies and in particular
implies that $H$ is invariant, that is, it is single valued.
It is also useful to represent the exchange symmetries of the
problem, e.g. the $2\leftrightarrow 3$ exchange, under which the 
Hamiltonian
is also invariant.
We have argued that the $U(2)$ symmetry will be replaced by a $U(N-1)$
symmetry in the nonrelativistic $N$-body case.

Given our understanding of the Hamiltonian structure, 
canonical quantization
is in principle straightforward, but the construction
of the canonical Hilbert space is not.
We have seen that in the 2-body case, it is possible to find
monodromic eigenfunctions by projecting over the $U(1)$
gauge variable $\theta$ from $-\infty$ to $+\infty$, after a weighted sum over monodromies and
careful regularization.
It is also possible that a similar sum and projection (for instance a harmonic projection
on $U(2)$ or on a proper subgroup) is able to define monodromic wave
functions for the 3-body system, together with their relevant
quantum numbers. Further analysis is needed in this direction.


\paragraph{Acknowledgments}
We acknowledge useful discussions with A. Cappelli and D. Seminara.
The work of SM is partly supported by the Agence Nationale 
de la Recherche (France),
contract ANR-06-JCJC-0084-02. 
He thanks the Department of Physics of the University
of Florence and the INFN
for hospitality and support at various stages of this work.


\newpage

\appendix

\section{\label{appB}$SU(2)$ monodromies and relabeling symmetry}

\subsection{Monodromies}

We compute the monodromies of the spinors $\sigma_a$
and $\sigma_b$ in Eq.~(\ref{defspinors})
and we check that they indeed transform according
to the matrices $M_{31}$ and $M_{32}$ in Eqs.~(\ref{M0}) and~(\ref{M1})
when the particles loop 
around each other.
While the transformation $\zeta\rightarrow e^{2i\pi}\zeta$ is
straightforward to perform on the expressions of  $f_{3b}$ and
$f_{2b}$ given in Eq.~(\ref{f}) and leads
to the diagonal matrix $M_{31}$ (see Eq.~(\ref{M0})),
the monodromy around the particle at position $\zeta=1$ 
is a bit more tricky.
The details being quite lengthy, we provide here only the main steps
in order for the reader to be able to reproduce the full calculation.

We start from the expressions of $f_{3b}$ and $f_{2b}$ in
Eq.~(\ref{f}), and we apply the well-known hypergeometric 
transformation
\cite{GR}
\begin{multline}
F(a,b,c,z)=\frac{\Gamma(c)\Gamma(c-a-b)}{\Gamma(c-a)\Gamma(c-b)}
F(a,b,a+b-c+1,1-z)\\
+\frac{\Gamma(c)\Gamma(a+b-c)}{\Gamma(a)\Gamma(b)}
(1-z)^{c-a-b}
F(c-a,c-b,1+c-a-b,1-z)
\label{eq:transfozto1-z}
\end{multline}
in order to change the argument of the latter from $\zeta$ to
$1-\zeta$.
Then, since the obtained 
hypergeometric functions are analytic around 
the point $1-\zeta=0$,
the monodromy transformations may be read from the prefactors.
We find
\begin{multline}
f_{3b}=\zeta^{\mu_1+\mu_3}\frac{\Gamma(\mu_1+\mu_2)\Gamma(\mu_1)}{\Gamma(\mu)}
F(\mu_1,1-\mu_2,1-\mu_2-\mu_3,1-\zeta)\\
+\zeta^{\mu_1+\mu_3}(1-\zeta)^{\mu_2+\mu_3}
\frac{\Gamma(1+\mu_3)\Gamma(-\mu_2-\mu_3)}{\Gamma(1-\mu_2)}
F(1+\mu_3,\mu,1+\mu_2+\mu_3,1-\zeta).
\end{multline}
As for $f_{2b}$, a similar transformation may be applied, which leads to
\begin{multline}
f_{2b}=\frac{\Gamma(\mu_1+\mu_3)\Gamma(1-\mu_1-\mu_3)\Gamma(\mu_2+\mu_3)}
{\Gamma(\mu)\Gamma(1-\mu_1)}F(-\mu_3,1-\mu,1-\mu_2-\mu_3,1-\zeta)\\
+\frac{\Gamma(\mu_1+\mu_3)\Gamma(\mu_2)\Gamma(1-\mu_1-\mu_3)\Gamma(-\mu_2-\mu_3)}
{\Gamma(-\mu_3)\Gamma(1-\mu)\Gamma(\mu)}(1-\zeta)^{\mu_2+\mu_3}
F(1-\mu_1,\mu_2,1+\mu_2+\mu_3,1-\zeta).
\end{multline}
Then, one applies a further transformation to the two hypergeometric
functions which appear in the previous expression,
\begin{equation}
\begin{split}
F(-\mu_3,1-\mu,1-\mu_2-\mu_3,1-\zeta)&=\zeta^{\mu_1+\mu_3}
F(1-\mu_2,\mu_1,1-\mu_2-\mu_3,1-\zeta)\\
F(1-\mu_1,\mu_2,1+\mu_2+\mu_3,1-\zeta)&=\zeta^{\mu_1+\mu_3}
F(\mu,1+\mu_3,1+\mu_2+\mu_3,1-\zeta)
\end{split}
\end{equation}
which enables one to write $f_{2b}$ in a form that is
similar to $f_{3b}$ as far as the hypergeometric functions are
concerned:
\begin{multline}
f_{2b}=\frac{s_1}{s_{13}}\frac{\Gamma(\mu_2+\mu_3)\Gamma(\mu_1)}
{\Gamma(\mu)}\zeta^{\mu_1+\mu_3}F(\mu_1,1-\mu_2,1-\mu_2-\mu_3,1-\zeta)\\
-\frac{s_{123}s_3}{s_{13}s_2}\frac{\Gamma(1+\mu_3)\Gamma(-\mu_2-\mu_3)}{\Gamma(1-\mu_2)}
\zeta^{\mu_1+\mu_3}(1-\zeta)^{\mu_2+\mu_3}
F(1+\mu_3,\mu,1+\mu_2+\mu_3,1-\zeta).
\end{multline}
One now recognizes that the hypergeometric functions which appear in
the transformed expressions for the $f_{3b}$ and $f_{2b}$ are the
same.

In this form, the hypergeometric functions are invariant by
the monodromy transformation $\zeta-1\rightarrow e^{2i\pi}(\zeta-1)$.
Only the prefactors transform, and in a trivial way.
Lengthy but straightforward calculations
lead to an expression of the relationship between the
transformed of the vector
$(f_{3b},f_{2b})$ and its untransformed form through the
multiplication by a matrix. Setting the relative normalizations
in front of $f_{3b}$ and $f_{2b}$ as in Eq.~(\ref{defspinors}),
we find that this matrix is precisely $M_{32}$ given in Eq.~(\ref{M0}).
Thus $\sigma_b$ transforms as a $SU(2)$ spinor 
(up to a $U(1)$ phase) under monodromy transformations.

\subsection{Label exchange symmetry}

We start from Eq.~(\ref{eq:Pwithsigma}) for 
the momentum
$P=(P_3,P_2)$, namely
\begin{equation}
P=Dz_2^{\mu-1}\left[\sigma_b+(\zeta_A-\zeta)\sigma_a\right].
\end{equation}
The dilation factor $D$ is obviously invariant under any
relabeling, see its definition in Eq.~(\ref{dilation0}).

Let us first exchange the labels of particles \#1 and \#3.
This amounts to replacing
$p_3$ by $p_1=-p_2-p_3$, to performing
the conformal transformation $\zeta\rightarrow\zeta/(\zeta-1)$
and the substitution $z_2\rightarrow z_2(1-\zeta)$.
The transformed momenta read
\begin{equation}
\tilde P=Dz_2^{\mu-1}(1-\zeta)^{\mu-1}\left(\tilde\sigma_b
+\frac{\zeta_A}{1-\zeta}\tilde\sigma_a\right),
\label{Ptilde310}
\end{equation}
where we have put a ``tilde'' sign above the transformed quantities.

We need to understand how $\sigma_b$ and $\sigma_a$ transform.
The normalization factors $\sqrt{N_2}$ 
and $\sqrt{N_3}$ are invariant in this case,
as seen from their definitions~(\ref{N}).
As for the transformations of the various functions $f$
defined in Eq.~(\ref{f}),
the key formula is the following identity between
hypergeometric functions:
\begin{equation}
F(a,b,c,\zeta/(\zeta-1))=(1-\zeta)^b
F(c-a,b,c,\zeta).
\end{equation}
Using this formula, 
expressing
$\tilde\sigma_{a}$ is then straightforward.
Defining
\begin{equation}
\tau_{31}\equiv\left(\begin{matrix}
-e^{i\pi\mu_{13}} & 0\\
0 & 1
\end{matrix}\right),
\end{equation}
we obtain
\begin{equation}
\tilde\sigma_a=(1-\zeta)^{2-\mu}\tau_{31}\cdot\sigma_a.
\label{eq:tildesigmaa}
\end{equation}
As for the transformation of the components
$f_{3b}$ and $f_{2b}$ of $\sigma_b$, one needs to use
of the respective contiguity relations for hypergeometric
functions
\begin{equation}
(c-1)F(a,b,c-1,z)-b F(a,b+1,c,z)+(b-c+1)F(a,b,c,z)=0
\end{equation}
with $a=1-\mu_2$, $b=\mu_1$, $c=1+\mu_{13}$, $z=\zeta$, and
\begin{equation}
c F(a,b,c,z)-b z F(a,b+1,c+1,z)-c F(a-1,b,c,z)=0
\end{equation}
with $a=1-\mu_3$, $b=1-\mu$, $c=1-\mu_{13}$, $z=\zeta$.
We arrive at
\begin{equation}
\tilde\sigma_b=(1-\zeta)^{1-\mu}\tau_{31}\cdot(\sigma_b-\zeta\sigma_a).
\label{eq:tildesigmab}
\end{equation}
Combining Eqs.~(\ref{eq:tildesigmaa}) 
and~(\ref{eq:tildesigmab})
with Eq.~(\ref{Ptilde310}), one easily finds
\begin{equation}
\tilde P=\tau_{31}\cdot P.
\end{equation}
We recognize that $\tau_{31}$
is the transformation matrix~(\ref{eq:P31tilde}).
(Note that the phases stem from the replacement of
$\zeta-1$ by $e^{-i\pi}(1-\zeta)$. The convention with the opposite
sign for the phase would have led to a different transformation matrix).

Next, we exchange the labels of
particles \#2 and \#3.
This amounts to exchanging $p_2$ and $p_3$,
to transforming
$\zeta$ into $1/\zeta$ and
to substituting $z_2$ by $z_2\zeta$.
The transformed momenta thus read
\begin{equation}
\tilde P=Dz_2^{\mu-1}\zeta^{\mu-1}
\left(
\tilde\sigma_b+\frac{\zeta_A-1}{\zeta}\tilde\sigma_a
\right).
\label{Ptilde0}
\end{equation}
The transformations of $\sqrt{N_2}$ and $\sqrt{N_3}$
are straightforward from their definitions~(\ref{N}).
Again, expressing the various components
$\tilde f$ of $\tilde\sigma$ with the help 
of the components of $\sigma$
requires involved manipulations of the hypergeometric functions.
We first need to use the formula
\begin{multline}
F(a,b,c,1/z)=\frac{\Gamma(b-a)\Gamma(c)}{\Gamma(b)\Gamma(c-a)}
\left(e^{-i\pi}z\right)^a F(a,a-c+1,a-b+1,z)\\
+\frac{\Gamma(a-b)\Gamma(c)}{\Gamma(a)\Gamma(c-b)}
\left(e^{-i\pi}z\right)^b F(b,b-c+1,b-a+1,z)
\end{multline}
in order to express the $\tilde f$'s, which are
hypergeometric functions of argument $1/\zeta$,
as linear combinations of the $f$'s.
Without any further transformation, we arrive at
\begin{equation}
\tilde\sigma_a=\zeta^{2-\mu}\tau_{32}\cdot\sigma_a,
\label{eq:tildesigmaaf}
\end{equation}
with
\begin{equation}
\tau_{32}\equiv\frac{1}{\sqrt{s_{12}s_{13}}}
\left(\begin{matrix}
e^{-i\pi\mu_1}\sqrt{s_2s_3} & e^{i\pi\mu_3}\sqrt{s_1s_{123}}\\
e^{i\pi\mu_2}\sqrt{s_1s_{123}} & -e^{i\pi\mu}\sqrt{s_2s_3}
\end{matrix}\right).
\end{equation}
The functions $\tilde f_{3b}$, $\tilde f_{2b}$
are deduced from the functions $\tilde f_{3a}$, $\tilde f_{2a}$
by simply replacing $\mu_2$ by $\mu_2+1$.
The hypergeometric functions $F(-\mu_2,\mu_1,\mu_{13},\zeta)$
and $F(1-\mu,1-\mu_3,2-\mu_{13},\zeta)$ then appear,
which are not part of our original basis $f_3,f_2$.
But they can actually 
be expressed as linear combinations of the 
$f$-functions using contiguity relations.
Thanks to the identity
\begin{equation}
c F(a,b,c,z)+(a-c)z F(a,b+1,c+1,z)+(z-1)c F(a,b+1,c)=0
\end{equation}
with $a=\mu_1$, $b=-\mu_2$, $c=\mu_{13}$, $z=\zeta$,
we write
\begin{equation}
F(\mu_1,-\mu_2,\mu_{13},z)=
\frac{\Gamma(\mu_{13})}{\Gamma(\mu_1)\Gamma(\mu_3)}
\zeta^{1-\mu_{13}}
\left[f_{3b}+(1-\zeta)f_{3a}\right].
\end{equation}
Using
\begin{multline}
(a-b)(a-c+1)F(a,b+1,c,z)+a(a-b)(z-1)F(a+1,b+1,c,z)\\
+(c-1)(a-b)F(a,b,c-1,z)=0
\end{multline}
with $a=1-\mu$, $b=-\mu_3$, $c=2-\mu_{13}$, $z=\zeta$,
we get
\begin{equation}
F(1-\mu,1-\mu_3,2-\mu_{13},\zeta)
=-\frac{\Gamma(\mu)}{\Gamma(1+\mu_2)\Gamma(\mu_{13}-1)}
\left[f_{2b}-(1-\zeta)f_{2a}\right].
\end{equation}
Inserting these identities in $\tilde \sigma_{b}$
expressed with the help of the hypergeometric
functions of argument $\zeta$, we find
\begin{equation}
\tilde\sigma_b=\zeta^{1-\mu}\tau_{32}\cdot
\left[
\sigma_b+(1-\zeta)\sigma_a
\right].
\label{eq:tildesigmabf}
\end{equation}
The replacement of 
$\tilde\sigma_a$ and $\tilde\sigma_b$ 
expressed with the help of $\sigma_a$ and $\sigma_b$
(Eq.~(\ref{eq:tildesigmaaf}) and~(\ref{eq:tildesigmabf}))
into Eq.~(\ref{Ptilde0})
leads to the transformation
\begin{equation}
\tilde P=\tau_{32}\cdot P,
\end{equation}
which is an other way to write 
Eq.~(\ref{eq:P32tilde}).


\end{document}